\documentclass[]{article}

\usepackage{amsmath, amstext, amsfonts}
\usepackage{amssymb}
\usepackage{multicol}
\usepackage[latin1]{inputenc}

\usepackage{graphicx}
\usepackage{dcolumn}
\usepackage{bm}

\begin{document}

\setcounter{MaxMatrixCols}{10}

\title{STATIONARY STATES IN A POTENTIAL WELL}

\maketitle

\vskip -4 ex
    {\sc 
  H.C. Rosu and  J.L. Mor\'an-L\'opez
  \medskip
  \footnotesize \vskip 1ex
       Instituto Potosino de Investigaci\'on Cient\'{\i}fica y Tecnol\'ogica,  SLP, Mexico

\medskip

 \vskip 0.5cm

\noindent {\bf Contents}\\
1. Introduction \\
2. Stationary Orbits in Old Quantum Mechanics  \\
2.1 Quantized Planetary Atomic Model   \\
2.2 Bohr's Hypotheses and Quantized Circular Orbits  \\
2.3 From Quantized Circles to Elliptical Orbits  \\
2.4 Experimental Proof of the Existence of Atomic Stationary States  \\
3. Stationary States in Wave Mechanics  \\
3.1 The Schr\"odinger Equation  \\
3.2 The Dynamical Phase \\
3.3 The Schr\"odinger Wave Stationarity  \\
3.4 Stationary Schr\"odinger States and Classical Orbits  \\
3.5 Stationary States as Sturm-Liouville Eigenfunctions \\
4. The Infinite Square Well: The Stationary States Most Resembling the Standing Waves on a String \\
5. 1D Parabolic Well: The Stationary States of the Quantum Harmonic
Oscillator\\
5.1 The Solution of the Schr\"odinger Equation  \\
5.2 The Normalization Constant  \\
5.3 Final Formulas for the HO Stationary States\\
5.4 The Algebraic Approach: Creation and Annihilation operators $\hat{a}^{\dagger}$ and $\hat{a}$  \\
5.5 HO Spectrum Obtained from Wilson-Sommerfeld Quantization
Condition\\
6. The 3D Coulomb Well: The Stationary States of the Hydrogen Atom \\
6.1 Separation of Variables in Spherical Coordinates \\
6.2 The Angular Separation Constants as Quantum Numbers \\
6.2.1 The Azimuthal Solution and the Magnetic Quantum Number \\
6.2.2 The Polar Solution and the Orbital Quantum Number \\
6.2.3 Space Quantization \\
6.3 Polar and Azimuthal Solutions Set Together \\
6.4 The Radial Solution and the Principal Quantum Number \\
6.5 Final Formulas for the Hydrogen Atom Stationary States  \\
6.6 Electronic Probability Density \\
6.7 Other 3D Coordinate Systems Allowing Separation of Variables \\
7. The 3D Parabolic Well: The Stationary States of the Isotropic Harmonic Oscillator \\
8. Stationary Bound States in the Continuum \\
9. Conclusions\\
Bibliography                                \\

Glossary

Bohr hypotheses: Set of hypotheses that Bohr introduced to explain
the stability of the atom.

Coulomb potential: Potential that holds the electrons attached to
the nucleus.

Creation and Annihilation operators: Mathematical objects that
create or annihilate particles.

Dynamical phase: It is the factor that contains the time dependence
of the wave function.

Hydrogen atom: The simplest atom in nature, consisting of one
electron revolving around a proton.

Laplace operator: A partial differential operator that contains the
second partial derivatives with respect to the space coordinates.

Magnetic quantum  number: A quantum number that is associated to the
direction of projection of the of the angular momentum

Orbital quantum number: The number associated to the quantization of
the orbital motion

Principal quantum number:  The number associated to the radial
solution of the Schr\"odinger equation and defines the energy of the
allowed orbits.

Quantized electron orbits: The electrons can move in the atom only
in very specific orbits.

Quantum Harmonic oscillator: The quantum analogous model for a
potential whose restoring force is proportional to the displacement.

Quantum Mechanics: Theory of the laws that rule the microscopic
world, i.e. at the atomic sizes and less.

Rutherford planetary atomic model: model to describe the atom in
which the nucleus is at the center and the electrons move around it.

Schr\"odinger equation: The differential equation that describes the
movement of atomic particles.

Square well potential:  A one-dimensional model of a potential with
vertical walls

Standing wave: The solutions to the Schr\"odinger equation that are
stationary.

Zeeman effect: The effect of splitting the electronic energy levels
when an atom is immersed in a magnetic field.

\vspace*{10pt}
\section*{Summary}

\noindent In the early days of the 20th century a set of important
observations in atomic and molecular physics could not be explained
on the basis of the laws of classical physics. One of the main
findings was the emission of light by excited atoms with very
particular frequencies. To explain those findings a new development
in physics was necessary, now known as quantum mechanics. In
particular, the concept of stationary states was introduced by Niels
Bohr, in 1913, in order to explain those observations and the
stability of atoms. According to E.C. Kemble (1929), the existence
of discrete atomic and molecular energy levels brought into
mechanics a new kind of atomicity superposed on the atomicity of
electrons and protons. We review here in a historical context the
topic of stationary states in the quantum world, including the
generalization to the primary ideas. We also discuss the stationary
states in one dimensional parabolic wells and the three dimensional
Coulomb and parabolic cases.

\section{Introduction}

At the beginning of the 20th century, some experimental observations
in atomic and molecular physics were impossible to explain on the
bases of classical physics. It was necessary to introduce
revolutionary concepts that lead to the foundation of quantum
mechanics. In this context the concept of stationary states played
an essential role in the development of new ideas that started to
explain the atomic world.

\noindent In 1908 J.R. Rydberg and W. Ritz studied in detail the
spectra of the light emitted by excited atoms. They found that the
spectra consisted of a set of defined lines of particular
wavelengths. Furthermore, the set of spectroscopic lines were
dependent only on the atom under study. Through the so-called
combination principle they put the data in a most systematic form.
Their principle states that the frequency of a particular spectral
line can be expressed as a difference between some members of the
set of frequency lines.

\noindent These findings could not be explained by the accepted
atomic model at that time, proposed by J.J. Thomson, claiming that
the electrons were embedded in a positive charged cloud, whose
extent was determined by the atomic radius. That model could not
explain also the data obtained by H.W. Geiger and E. Mardsen, who
under the supervision of Rutherford, were studying the interaction
of charged $\alpha$-particles with gold foils \cite{Ruther11}. They
observed that a considerable fraction of the $\alpha$-particles was
deflected by large angles. This effect could not be attributed to
the electrons since they are much less massive. Thus, they concluded
that the source of deflection must be the positive charge
concentrated in a much smaller volume than the one generated by the
atomic radius. In 1911 Rutherford proposed a new atomic model which
assumed that all the positive charge is located at the center of the
atom with a very dense distribution with a radius much smaller that
the atomic one. The electrons then would circulate around the
nucleus in a similar way as the planets move around the sun.

\noindent Although Rutherford's planetary atomic model explained
qualitatively well the deflection of $\alpha$-particles, it had two
major deficiencies. First it could not account for the spectra of
radiation from atoms, which was not continuous but discrete. The
other major problem was that, according to electrodynamics, an
electron moving around the atom is under a constant acceleration,
must radiate energy. This fact would lead to a situation in which
the electron would loose energy continuously and would collapse with
the nucleus.

\section{Stationary Orbits in Old Quantum Mechanics}

\subsection{Quantized Planetary Atomic Model}

In 1911, the two-and-a half-thousand-year-old philosophical concept
of atom turned into a scientific matter when Rutherford's planetary
atomic model emerged from the interpretation of the experimental
data on the scattering of $\alpha$ particles \cite{Ruther11}. The
curious fact that has been noticed while these particles were shut
to gold foils was that some of them bounced as if they were
colliding with very massive objects. To explain these findings
Rutherford proposed that the atom was composed by a positive central
massive nucleus and the electrons were revolving around it, i.e.
very similar to a miniature solar system. However, this famous model
was not electrodynamically viable. Atomic stability was simply not
assured for Rutherford's semiempiric model, since accelerated
charges radiate energy and the electrons moving around the nucleus
would loss energy and eventually collapse with the nucleus.

Another important set of empirical set of data, is that obtained
from the emission of light by excited atoms. It was observed that
the light emitted had a very characteristic frequencies and was a
footprint for each atom. These observations were put in a systematic
form in 1908 through the so-called combination principle formulated
by J.R. Rydberg and W. Ritz. Their principle says that the frequency
of a spectral emission or absorption line can be expressed as a
difference between the members of a set of well defined frequency
terms. Rutherford's model was completely silent on the dynamical
origin of the spectral lines. It was the great merit of Bohr to
formulate in 1913 the hypotheses, or postulates, that could allow
the explanation of the atomic spectral lines based on the planetary
atomic structure.

\subsection{Bohr's Hypotheses and Quantized Circular Orbits}

The hypotheses that Bohr added to the Rutherford model in order to
explain the spectroscopic information are the following
\cite{Bohr1913}

\begin{quote}

\begin{enumerate}

\item  An atom can exist only in special states with discrete values
of energy. In other words, the electrons moving around an atom can
be found only in certain special orbits that Bohr called {\em
stationary states}.

\item When an atom makes a {\em transition} from one stationary
state to another, it emits or absorbs radiation whose frequency
$\nu$ is given by the frequency condition
\begin{equation}
h\nu =E_1-E_2~,
\end{equation}
where $E_1$ and $E_2$ are the energies of two stationary states.

\item In the stationary states, the electrons move according to the
laws of classical theory. However, only those motions are performed
for which the following {\em quantum condition} is fulfilled
\begin{equation}
\oint p\,dq=nh~, \qquad (n=1,2,3,...;)~,
\end{equation}
where $p$ is the momentum of the electron and $q$ is its coordinate
along the stationary orbit. The integration should be taken along
the orbit over one period of the cyclic motion.

\end{enumerate}

\end{quote}

Bohr's theory claimed that those frequency terms, when multiplied by
$h$, give distinct energy levels in which the electrons move around
the nucleus. This meant that these were the only possible states in
which the electrons in the atom could exist.

Let us assume that an electron in a Hydrogen atom is revolving
around the nucleus on a circular orbit according to the Newtonian
equations of motion. For a circular orbit, the absolute value of the
momentum $p$ is constant and then the quantum hypothesis (3) leads
to

\begin{equation}
p\cdot 2\pi a=nh~, \qquad (n=1,2,3,...)
\end{equation}

where $a$ is the radius of the orbit. Thus, $a$ is given by the
value of the momentum that can be obtained from the balance between
the centrifugal force and the Coulomb force, i.e.,

\begin{equation}
\frac{p^2}{ma}=\frac{e^2}{4\pi \epsilon _0a^2} \  .
\end{equation}

Combining the two equations, one obtains

\begin{equation}
a_n=\frac{\epsilon _0h^2n^2}{\pi m e^2} \qquad (n=1,2,3,...)~.
\end{equation}

The latter formula gives the radii of the quantized electron circles
in the hydrogen atom. In particular, $a_1\equiv a_B=\frac{\epsilon
_0h^2}{\pi m e^2}$, is known as the Bohr radius and is taken as an
atomic length unit.

\subsection{From Quantized Circles to Elliptical Orbits}

Wilson \cite{wilson15} and Sommerfeld \cite{sommer16} extended
Bohr's ideas to a large variety of atomic systems between 1915 and
1916.

The main idea is that the only classical orbits that are allowed as
stationary states are those for which the condition
\begin{equation}\label{ws1}
\oint p_k dq_k=n_k h \qquad k=1,...,n~,
\end{equation}

with $n_k$ a positive integer, is fulfilled. The weak theoretical
point is that in general these integrals can be calculated only for
{\em conditionally periodic systems}, because only in such cases a
set of coordinates can be found, each of which goes through a cycle
as a function of the time, independently of the others. Sometimes
the coordinates can be chosen in different ways, in which case the
shapes of the quantized orbits depend on the choice of the
coordinate system, but the energy values do not.

In particular, when the 3D polar coordinates are employed, Eq.~(6)
gives the Sommerfeld ellipses characterized by

\begin{equation}\label{ws2}
\oint p_r dr =n_r h~, \quad \oint p_\theta d\theta =n_\theta h~,
\quad \oint p_\phi d\phi =n_\phi h~.
\end{equation}

Now, since $p_\phi$ is a constant, one gets immediately the
`quantization' of the angular momentum of the ellipse along the $z$
axis

\begin{equation}\label{ws3}
p_\phi=\frac{mh}{2\pi}~, \quad m=\pm 1, \pm 2, \cdot \cdot \cdot ~.
\end{equation}

The quantum number $m$ was called the magnetic quantum number by
Sommerfeld who used it as a measure of the direction of the orbit
with respect to the magnetic field and thus explaining the Zeeman
effect, i.e., the splitting of the spectroscopic lines in a magnetic
field. Unless for the value $m=0$ which is considered as unphysical,
this `old' $m$ is practically equivalent with Schr\"odinger's $m$,
which mathematically is the azimuthal separation constant but has a
similar interpretation.

Interestingly, and this is sometimes a source of confusion, the
`old' azimuthal quantum number is denoted by $k$ and is the sum of
$n_\theta$ and $m$. It gives the shape of the elliptic orbit
according to the relationship $\frac{a}{b}=\frac{n}{k}$, where
$n=n_r+k$, established by Sommerfeld. Actually, this $k$ is
equivalent to Schr\"odinger's orbital number $l$ plus 1, but again
their mathematical origin is quite different.

\begin{figure}
\centering
    \includegraphics[width=5.5in]{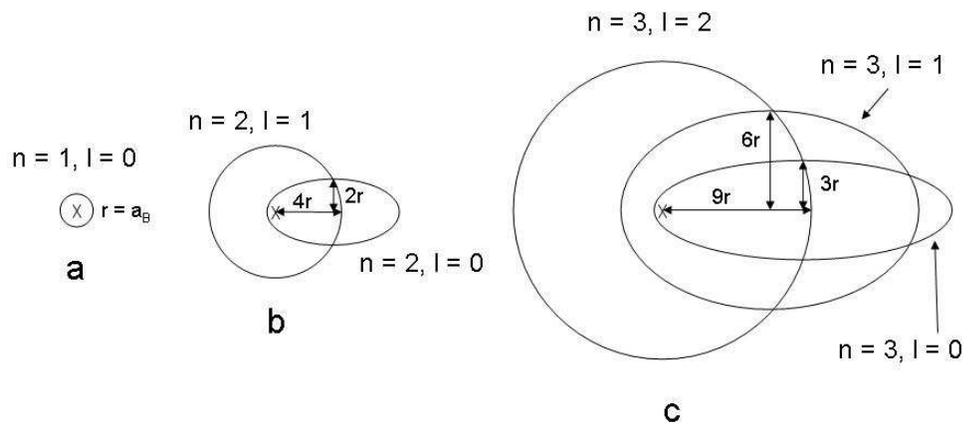}  
    \caption{Bohr-Sommerfeld electron orbits for $n=1$, $2$, and $3$, and the allowed values for $l$.}
    \label{BS-elipses}
\end{figure}

\begin{figure}
\centering
    \includegraphics[width=5.5in]{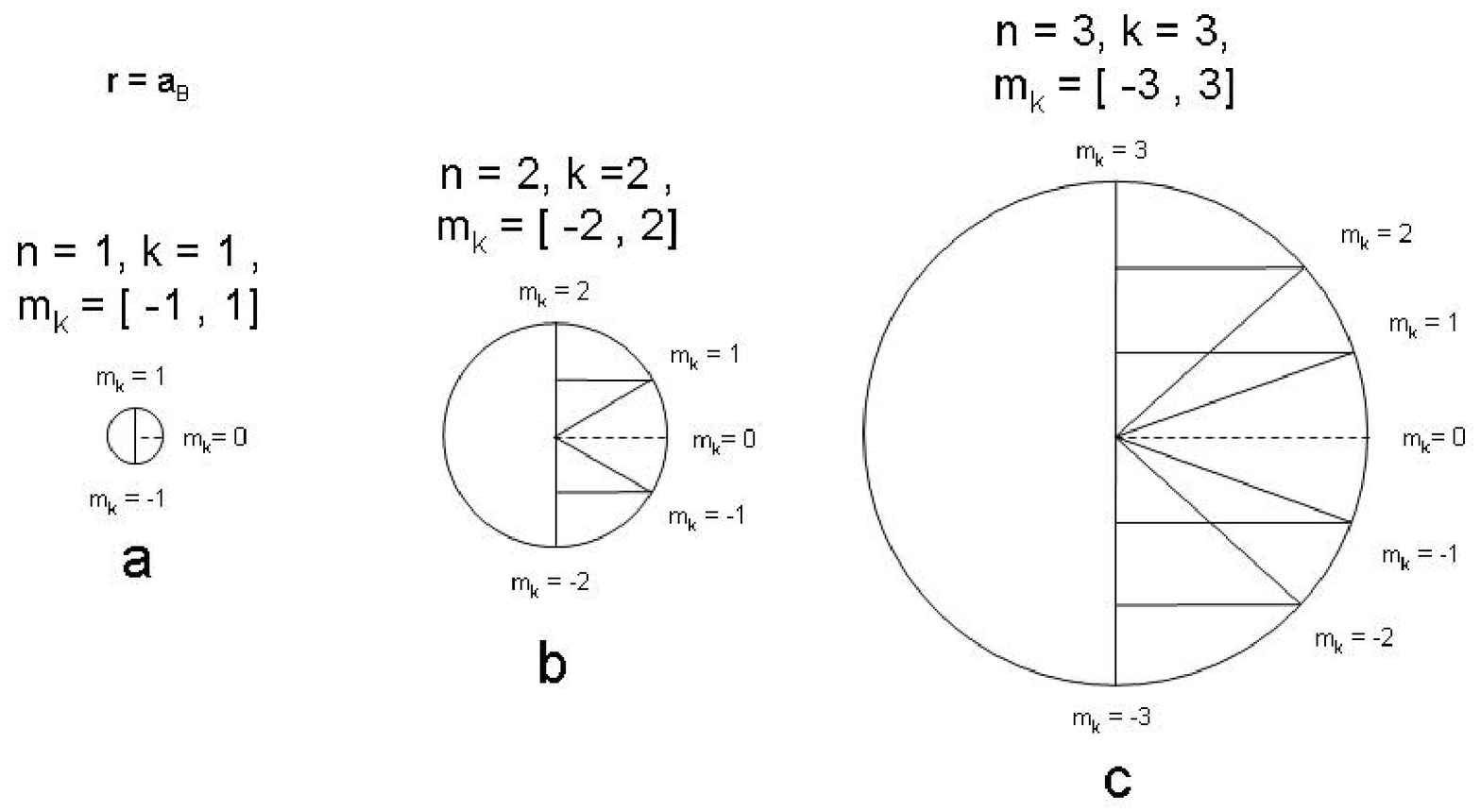}
    \caption{Spatial quantization of Bohr-Sommerfeld orbits for azimuthal numbers $k$ = 1,2, and 3.}
    \label{123}
\end{figure}

\subsection{Experimental Proof of the Existence of Atomic Stationary States}

The existence of discrete atomic energy levels was evidenced for the
first time by J. Franck and G. Hertz in 1914 \cite{FH14}. They
observed that when an electron collides with an atom (mercury in
their case), a transfer of a particular amount of energy occurred.
This energy transfer was recorded spectroscopically and confirmed
Bohr's hypotheses that atoms can absorb energy only in quantum
portions. Even today, the experiment is preferentially done either
with mercury or neon tubes. From the spectroscopic evidence, it is
known that the excited mercury vapor emits ultraviolet radiation
whose wavelength is 2536 \AA, corresponding to a photon energy
$h\nu$ equal to 4.89 eV.

The famous Franck-Hertz curves represent the electron current versus
the accelerating potential, shown in Fig. 3. The current shows a
series of equally spaced maxima (and minima) at the distance of
$\sim$ 4.9 V. The first dip corresponds to electrons that lose all
their kinetic energy after one inelastic collision with a mercury
atom, which is then promoted to its first excited state. The second
dip corresponds to those electrons that have the double amount of
kinetic energy and loses it through two inelastic collisions with
two mercury atoms, and so on. All these excited atoms emit the same
radiation at $\sim$ 2536 \AA. But which is the `first' excited state
of mercury? It is spectroscopically denoted by $^3P_1$ in
Fig.~(\ref{Mercury}). Notice that the other two $P$ states cannot
decay to the ground state $^1S_0$ because the dipole emission is
forbidden for them and therefore they are termed metastable. More
details, such that the observed peak separation depends on the
geometry of the tube and the Hg vapor pressure, are explained in the
readable paper of Hanne \cite{hanne}.

\begin{figure}
\centering
    \includegraphics[width=3.32in]{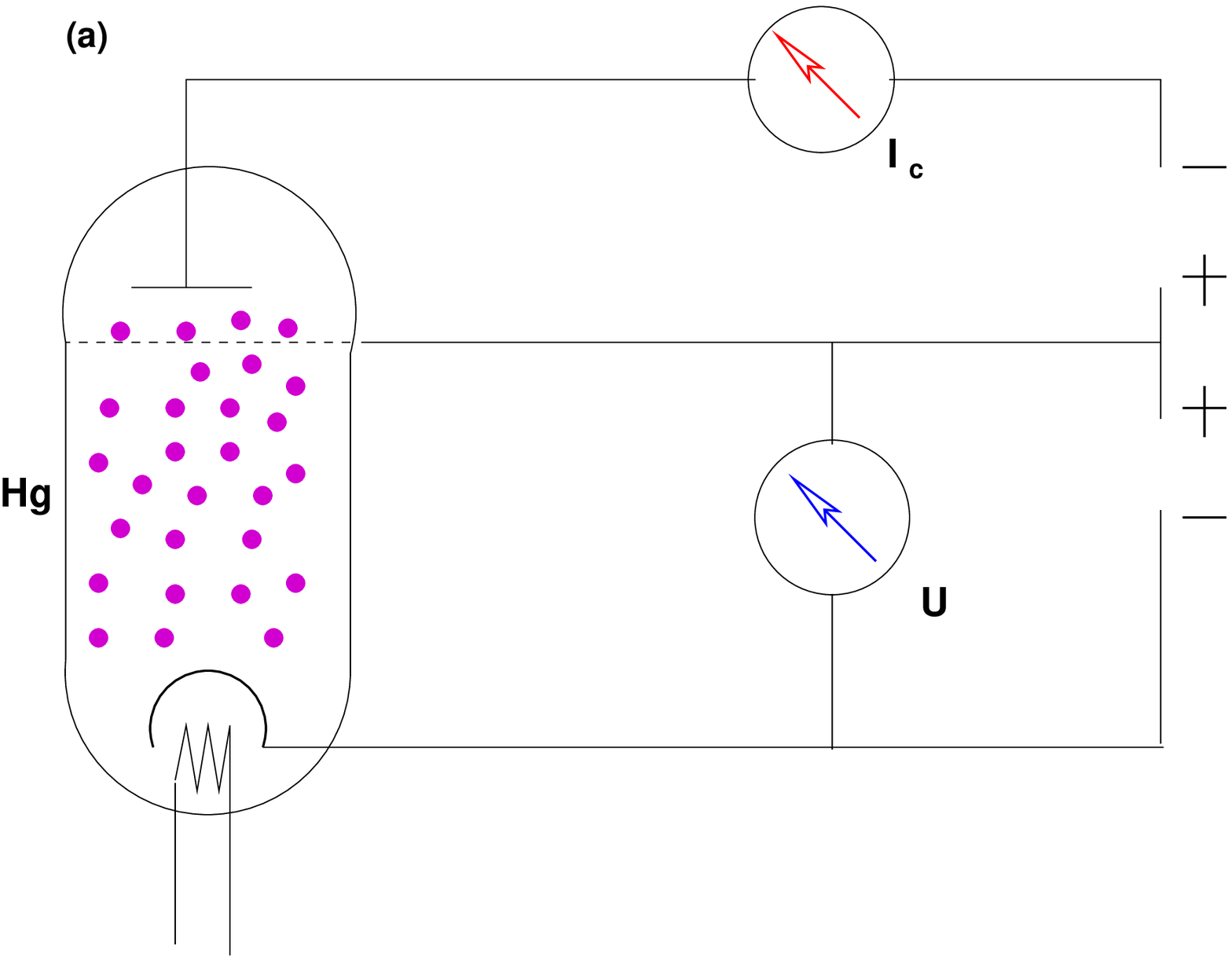}\qquad
    \includegraphics[width=3.32in]{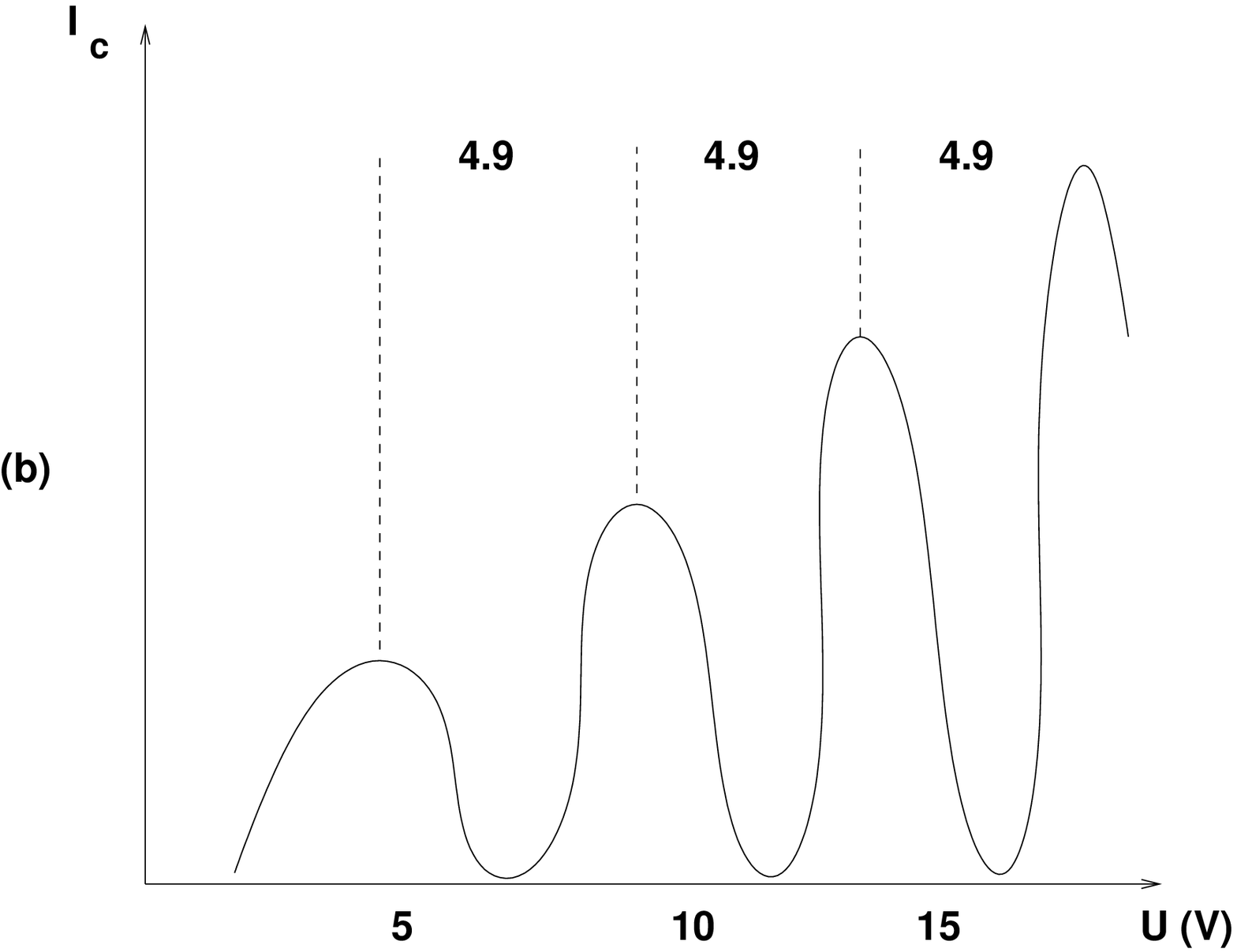}
    \caption{\small (a) Schematic diagram of the Franck-Hertz experiment, where the tube is filled with
a gas of Mercury ; (b) typical plot recorded in
    a Franck-Hertz experiment with mercury, showing the periodic maxima.}
    \label{FH-exp}
\end{figure}

\begin{figure}
\centering
    \includegraphics[width=2.8in]{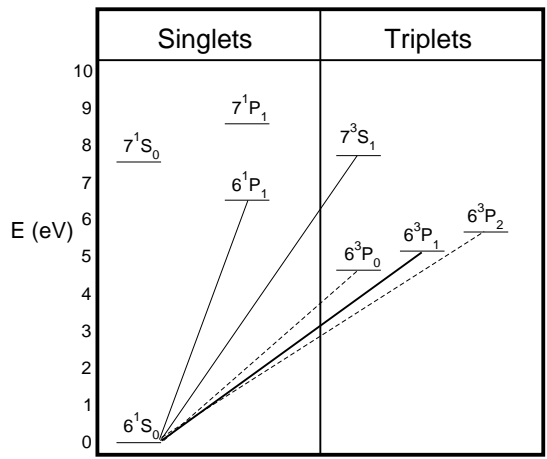}
    \caption{\small Simplified Mercury's level diagram in the low-energy region in which the so-called hyperfine structure is neglected.
    The numbers 6 and 7 are Bohr's `whole numbers' or Schr\"odinger's principal quantum numbers.}
    \label{Mercury}
\end{figure}

\section{Stationary States in Wave Mechanics}

\subsection{The Schr\"odinger Equation}

According to L. Pauling and E. Bright Wilson Jr. \cite{paul-bright},
already in the years 1920-1925 a decline of the `old quantum theory'
as the Bohr-Sommerfeld atomic theory is historically known and which
is based on the `whole number' quantization of cyclic orbits was
patent; only very recently there is some revival, especially in the
molecular context \cite{revB}. But in 1925, a quantum mechanics
based on the matrix calculus was developed by W. Heisenberg, M.
Born, and P. Jordan and the best was to come in 1926 when
Schr\"odinger in a series of four papers developed the most employed
form of quantum mechanics, known as {\em wave mechanics}. The
advantage of his theory of atomic motion is that it is based on
standard (partial) differential equations, more exactly on the
Sturm-Liouville theory of self-adjoint linear differential
operators. Schr\"odinger starts the first paper in the 1926 series
with the following sentence \cite{S26a}:

\begin{quote}
In this paper I wish to consider, first the simplest case of the
hydrogen atom, and show that the customary quantum conditions can be
replaced by another postulate, in which the notion of `whole
numbers', merely as such, is not introduced.
\end{quote}

Indeed, he could obtain the basic equation of motion in
nonrelativistic quantum mechanics, the  so called Schr\"odinger
equation for the wavefunctions $\Psi(x,t)$, and provided several
analytic applications, among which was the hydrogen atom. The
original derivation is based on the variational calculus within the
Sturm-Liouville approach and was given eighty years ago. In his
first paper of 1926, Schr\"odinger states that the wavefunctions
$\Psi$ should be such as to make the `Hamilton integral'
\begin{equation}
{\cal J}_{S}[\Psi]= \int \left(\hbar ^2 T(q, \partial \psi/\partial
q) +\Psi ^2 V\right)d\tau  \ ,
\end{equation}
stationary subject to the normalizing condition $\int \Psi ^2d\tau
=1$ which can be incorporated through the Lagrange multipliers
method. The Euler-Lagrange equation of the functional ${\cal
J}_{S}[\Psi]$ is the time-dependent Schr\"odinger equation
\begin{equation}
i\hbar \frac{\partial \Psi}{\partial t}= H\Psi~.
\end{equation}
When the wave function of the time-dependent Schr\"odinger equation
is written in the multiplicative form $\Psi(x,t)=\psi (x){\cal
F}(t)$ one obtains a complete separation of the space and time
behaviors of $\Psi$: on one side, one gets the stationary
Schr\"odinger equation for $\psi (x)$,
\begin{equation}
-\frac{h^2}{2m}\frac{d^2\psi}{dx^2}+V(x)\psi=E\psi ~,
\end{equation}
and on the other side, the simple time-dependent equation for the
logderivative of ${\cal F}$
\begin{equation}
i\hbar \frac{d\log {\cal F}}{dt}=E
\end{equation}

This decoupling of space and time components is possible whenever
the potential energy is independent of time.

The space component has the form of a standing-wave equation. Thus,
it is correct to regard the time-independent Schr\"odinger equation
as a wave equation from the point of view of the spatial
phenomenology.

\subsection{The Dynamical Phase}

Furthermore, the time-dependence is multiplicative and reduces to a
modulation of the phase of the spatial wave given by
\begin{equation}
{\cal F}=e^{-iEt/\hbar}=\cos(Et/\hbar)-i\sin(Et/\hbar)~.
\end{equation}
The phase factor ${\cal F}=e^{-iEt/\hbar}$ is known as the {\em
dynamical phase}. In recent times, other parametric phases have been
recognized to occur, e.g., the Berry phase. The dynamical phase is a
harmonic oscillation with angular frequency $\omega = E/h$ and
period $T=h/E$. In other words, a Schr\"odinger wavefunction is
flickering from positive through imaginary to negative amplitudes
with a frequency proportional to the energy. Although it is a wave
of constant energy it is not stationary because its phase is time
dependent (periodic). However, a remarkable fact is that the product
$\Psi ^{*}\Psi$, i.e., the modulus $|\Psi|^2$ of the Schr\"odinger
constant-energy waves remains constant in time
\begin{equation}
\Psi ^{*}\Psi = \psi ^{*}\psi
\end{equation}
It is in the sense of their constant modulus that Schr\"odinger
constant-energy waves are called stationary states.

\subsection{The Schr\"odinger Wave Stationarity}

Thus, non-relativistic quantum stationarity refers to waves of
constant energy and constant modulus, but not of constant phase,
which can occur as solutions of Schr\"odinger equation for
time-independent potentials. In the Schr\"odinger framework, the
dynamical systems are usually assumed to exist in stationary states
(or waves of this type). It is worth noting that the preferred
terminology is that of states and not of waves. This is due to the
fact that being of constant energy the Schr\"odinger stationary
waves describe physical systems in configurations (or states) of
constant energy which can therefore be naturally associated to the
traditional conservative Hamiltonian systems. Moreover, the
localization of these waves can be achieved by imposing appropriate
boundary conditions.

\subsection{Stationary Schr\"odinger States and Classical Orbits}

In the Schr\"odinger theory, a single stationary state does not
correspond to a classical orbit. This is where the Schr\"odinger
energy waves differ the most from Bohr's theory which is based on
quantized classical cyclic trajectories. To build a wave entity
closer to the concept of a classical orbit, one should use
superpositions of many stationary states, including their time
dependence, i.e., what is known as wave packets. Only monochromatic
plane waves of angular frequency $\omega$ corresponds through the
basic formula $E=\hbar \omega$ to a well-defined energy $E$ of the
`classical' particle but unfortunately there is no relationship
between the wavevector $k$ and the momentum $p$ of the corresponding
particle since a plane wave means only the propagation at constant
(phase) velocity of infinite planes of equal phase. In other words,
a criterium for localization is required in order to define a
classical particle by means of a wave approach.

In the one-dimensional case, a wave packet is constructed as follows
\begin{equation}\label{wp1}
\psi(x,t)=\int _{-\infty}^{+\infty}f(k')e^{i(k' x-\omega ' t)} dk' \
,
\end{equation}
with obvious generalization to more dimensions. If $f(k')$ is
written in the polar form $F(k')e^{i\alpha}$ and $F$ is chosen with
a pronounced peak in a wavenumber region of extension $\Delta k$
around the point $k'=k$, then the wave packet is localized in a
spatial region of extension $\Delta x\approx \frac{1}{\Delta k}$
surrounding the ``center of the wavepacket". The latter is
equivalent to the concept of material point in classical mechanics
and travels uniformly with the group velocity
$v_g=\frac{d\omega}{dk}$. This is the velocity that can be
identified with the particle velocity $v=\frac{dE}{dp}$ in classical
mechanics and which leads to the de Broglie formula $p=\hbar
k=\frac{h}{\lambda}$.

\subsection{Stationary States as Sturm-Liouville Eigenfunctions}

The mathematical basis of Schr\"odinger wave mechanics is the
Sturm-Liouville (SL) theory of self-adjoint linear differential
operators established in the 19th century, more specifically the SL
eigenvalue problem, which is to find solutions of a differential
equation of the form
\begin{equation}\label{sl1}
{\cal L}u_n(x)\equiv
\frac{d}{dx}\left[p(x)\frac{du_n(x)}{dx}\right]-q(x)u_n(x)=-\mu _n
w(x)u_n(x)
\end{equation}
subject to specified boundary conditions at the beginning and end of
some interval $(a,b)$ on the real line, e.g.
\begin{equation}\label{sl2}
Au(a)+Bu'(a)=0 \end{equation}
\begin{equation}\label{sl3}
Cu(b)+Du'(b)=0 \end{equation} where $A,B,C,D$ are given constants.
The differential operator ${\cal L}$ in (\ref{sl1}) is rather
general since any second order linear differential operator can be
put in this form after multiplication by a suitable factor. The
boundary conditions are also rather general including the well-known
Dirichlet and Neumann boundary conditions as particular cases but
other possibilities such as periodic boundary conditions
\begin{equation}\label{sl4}
u(x)=u(x+b-a)
\end{equation}
could be of interest in some cases, especially for angular
variables.

\medskip

The SL eigenvalue problem is an infinite dimensional generalization
of the finite dimensional matrix eigenvalue problem
\begin{equation}\label{sl5}
Mu=\mu u
\end{equation}
with $M$ an $n\times n$ matrix and $u$ an $n$ dimensional column
vector. As in the matrix case, the SL eigenvalue problem will have
solutions only for certain values of the eigenvalue $\mu _n$. The
solutions $u_n$ corresponding to these $\mu _n$ are the
eigenvectors. For the finite dimensional case with an $n\times n$
matrix $M$ there can be at most $n$ linearly independent
eigenvectors. For the SL case there will in general be an infinite
set of eigenvalues $\mu _n$ with corresponding eigenfunctions
$u_n(x)$.

\medskip

The differential equations derived by separating variables are in
general of the SL form, the separation constants being the
eigenvalue parameters $\mu$. The boundary conditions
(\ref{sl2},\ref{sl3}) are determined by the physical application
under study.

\medskip

The solutions $u_n(x),\, \mu _n$ of a SL eigenvalue problem have
some general properties of basic importance in wave (quantum)
mechanics.

\medskip

If $u(x)$ and $v(x)$ are arbitrary twice differentiable solutions of
a SL operator, then by integrating by parts
\begin{equation}\label{slosc10}
\int _a^b dx[v{\cal L}u-u{\cal L}v]=
p\left(v\frac{du}{dx}-u\frac{dv}{dx}\right)\bigg |_a^b~.
\end{equation}
An operator ${\cal L}$ which satisfies (\ref{slosc10}) is said to be
self-adjoint. Any second order linear differential operator can be
put in this self adjoint form by multiplication by a suitable
factor.

\medskip

It is easy to show that for functions $u(x)$, $v(x)$ satisfying
boundary conditions of the standard SL form or the periodic boundary
condition (\ref{sl4}) the right hand side of (\ref{slosc10})
vanishes. For both of this case we then have
\begin{equation}\label{sl11}
\int _a^b dxv{\cal L}u=\int _a^b dxu{\cal L}v~.
\end{equation}


Consider now two different eigenfunctions $u_n(x)$, $u_m(x)$
belonging to different eigenvalues $\lambda _n\neq \lambda _m$:
\begin{eqnarray}\label{sl12}
{\cal L}u_n(x)&=&-\mu _n w (x) u_n(x)~,\\
{\cal L}u_m(x)&=&-\mu _m w (x) u_m(x)~.
\end{eqnarray}
Multiplying the first equation by $u_m$ and the second one by $u_n$,
integrating and subtracting, we find:
\begin{equation}\label{sl13}
\int _a^b dx[u_m{\cal L}u_n-u_nLu_m]=-(\mu _n-\mu _m)\int _a^b dx
w(x)u_n(x)u_m(x)~.
\end{equation}
The left hand side will vanish for either set of boundary conditions
we consider here, so for either of these cases, we find the
orthogonality condition

\medskip

\begin{equation}\label{sl14}
\int _a^b dx w(x)u_n(x)u_m(x)=0~, \qquad \mu _n \neq \mu _m~.
\end{equation}
Two functions $u_n(x)$, $u_m(x)$ satisfying this condition are said
to be orthogonal with weight $w(x)$. Moreover, if $w(x)$ is non
negative, one can introduce the SL normalization of the $u_n$ as
follows:

\medskip

\begin{equation}\label{sl15}
\int _a^b dx w(x)[u_n(x)]^2=1~.
\end{equation}

The most important property of the eigenfunctions of a SL problem is
that they form a complete set. This means that an arbitrary function
$\psi(x)$ can be expanded in an infinite series of the form
\begin{equation}\label{slcompl}
\psi(x)=\sum _n a_nu_n(x)~.
\end{equation}
The expansion coefficients $a_n$ are determined by multiplying
(\ref{slcompl}) by $w(x)u_n(x)$, integrating term by term, and using
the orthogonality relation
\begin{equation}\label{slcompl-1}
a_n=\frac{\int _a^b dx w(x) u_n(x)\psi(x)}{\int _a^bdx
w(x)[u_n(x)]^2}~.
\end{equation}

According to Courant and Hilbert \cite{ch53}, every piecewise
continuous function defined in some domain with a square-integrable
first derivative may be expanded in an eigenfunction series which
converges absolutely and uniformly in all subdomains free of points
of discontinuity; at the points of discontinuity it represents (like
the Fourier series) the arithmetic mean of the right and left hand
limits. (This theorem does not require that the functions expanded
satisfy the boundary conditions.)

\section{The Infinite Square Well: The Stationary States Most Resembling the Standing
Waves on a String}

We have already commented that the time-independent Schr\"odinger
equation  has the form of a standing-wave equation. This is a very
instructive analogy and allows to obtain the correct energy values
for the case of the infinite square well using only the de Broglie
wave concept without even introducing the Schr\"odinger equation
\cite{cerny86}.

We remind the treatment of the string standing waves in the case of
a finite homogeneous string of total length $L$. Provided that the
origin of the coordinate system is placed in its center and the $x$
direction is chosen parallel to it, the space dependent part of its
standing waves is given by
\begin{equation}\label{l6}
u(x) = A\cos kx +B \sin kx
\end{equation}
where $x\in (-L/2, L/2)$ and $A$, $B$, and $k$ are constants.
Imposing the usual (Dirichlet) boundary conditions
\begin{equation}\label{l7}
u(-L/2)=u(L/2)=0 \ ,
\end{equation}
equation (\ref{l6}) takes the form
\begin{equation}\label{l8}
u_n(x) = \left \{
\begin{array}{ll}
A\cos k_nx~, & \mbox{if $n> 0$ and odd}\\
B\sin k_nx ~, & \mbox{if $n> 0$ and even}
\end{array}
\right.
\end{equation}
where
\begin{equation}\label{l9}
x\in ( -L/2,L/2) \end{equation} and
\begin{equation}\label{l10}
k_n=\frac{n\pi}{L}~, \qquad (n=1,2,3,...)~.
\end{equation}

These functions - the normal modes of the string under consideration
- form a complete set with respect to physically reasonable
functions defined within the interval (\ref{l9}) and satisfying
equation (\ref{l7}), i.e., any such function $U(x)$ can be written
as the Fourier series
\begin{equation}\label{l11}
U(x)=\sum _{n=1}^{\infty}c_n u_n(x)~.
\end{equation}

\medskip

We are ready now to state the analogy, which is based on the
following steps.

\begin{itemize}

\item A standing de Broglie wave corresponds to a quantum particle strictly confined
to the region $-L/2\leq x\leq L/2$, i.e., in an infinite square-well
potential shown in (Fig.~(\ref{fig:Iwell})
\begin{equation}\label{isw1}
V(x) = \left \{
\begin{array}{ll}
0~, & \mbox{if $|x|\leq L/2$}\\
-\infty ~, & \mbox{if $|x| > L/2~.$}
\end{array}
\right.
\end{equation}

\begin{figure} 
\centering
    \includegraphics[width=3.2in]{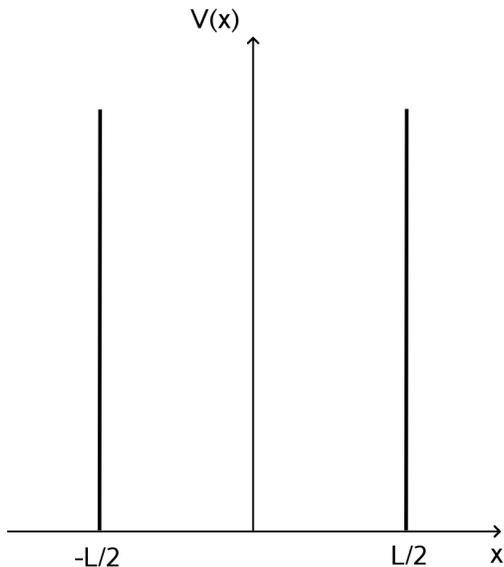}
    \caption{The infinite square well potential. This case is mathematically the most analogous to the
    classical string standing waves.}
    \label{fig:Iwell}
\end{figure}

Because of the form of this potential, it is assumed that there is
no asymptotic tail of the wave functions in the outside regions to
the well, i.e., $\psi (x)=0$ for $|x|>L/2$. In physical terms, that
means that the particle is completely localized within the wall and
the boundary conditions are of the Dirichlet type.

In the interior region, the Schr\"odinger equation is the simplest
possible:
\begin{equation}\label{isw0}
-\frac{h^2}{2m}\frac{d^2\psi}{dx^2}-E\psi=0 \ ,
\end{equation}

whose solutions (energy eigenfunctions) are:
\begin{equation}\label{isw2}
\phi _n(x) =\left \{
\begin{array}{lll}
(2/L)^{1/2}\sin(n\pi x/L)~, & n \, \mbox{even}~,  \quad
\mbox{$|x|\leq
L/2$} \ , \\
(2/L)^{1/2}\cos(n\pi x/L)~, & n \, \mbox{odd}~, \quad \mbox{$|x|\leq
L/2$} \ , \\
0 ~, & \quad  \mbox{$|x| > L/2~,$}
\end{array}
\right.
\end{equation}
and the energy eigenvalues are:
\begin{equation}\label{isw3}
E_n=\frac{(\hbar k_n)^2}{2m}=\frac{\hbar ^2 \pi ^2 n^2}{2mL^2}~.
\end{equation}

\item The amplitude of this standing wave in the $n$th stationary
state is proportional to $\sin (n\pi z/L)$. This corresponds
strictly to the analogy with standing waves on a classical string.

\item The $n$th standing wave is presented as a superposition of two
running waves. The wave travelling right, with wavelength $\lambda
_n=2L/n$ has de Broglie momentum $\hbar \pi n/L$ and the left
travelling wave has opposed de Broglie momentum $-\hbar \pi n/L$.
The resulting energy is then quantized and given by
\begin{equation}
E_n=\frac{p_n^2}{2m}=\frac{\hbar ^2 \pi ^2 n^2}{2mL^2}~.
\end{equation}
\end{itemize}

Because of the extremely strong confinement of the infinite square
well it seems that this case is only of academic interest. However,
two-dimensional strong confinement of electrons by rings of adatoms
(corrals) have been reported in the literature \cite{crommie93}.

\newcommand{\bc}{\begin{center}}
\newcommand{\ec}{\end{center}}
\newcommand{\dd}{\dagger}
\newcommand{\ad}{a^{\dd}}
\newcommand{\m}{\mid}
\newcommand{\ii}{\'{\i}}

\section{\Large 1D Parabolic Well: The Stationary States of the Quantum Harmonic Oscillator}

\subsection{The Solution of the Schr\"odinger Equation}

The harmonic oscillator (HO) is one of the fundamental paradigms of
Physics. Its utility resides in its simplicity which is manifest in
many areas from classical physics to
quantum electrodynamics and theories of gravitational collapse.\\
It is well known that within classical mechanics many complicated
potentials can be well approximated close to their equilibrium
positions $a_i$ by HO potentials as follows
\begin{equation} \label{ho1}
V(x) \sim \frac{1}{2}V^{\prime\prime}(a_i)(x-a_i)^2~.
\end{equation}

For this case, the classical Hamiltonian function of a particle of
mass {\em m}, oscillating at the frequency $\omega$ has the
following form:
\begin{equation}\label{ho2}
H=\frac{p^2}{2m}+\frac{1}{2}m\omega^2x^2 \ ,
\end{equation}
and the quantum Hamiltonian corresponding to the configurational
space is given by

\begin{equation}\label{ho4}
\hat{H}=\frac{1}{2m}\left(-i\hbar\frac{d}{dx}\right)^2
+\frac{1}{2}m\omega^2x^2=-\frac{\hbar^2}{2m}\frac{d^2}{dx^2}+\frac{1}{2}m\omega^2x^2~.
\end{equation}

\begin{figure} 
\centering
    \includegraphics[width=3.5in]{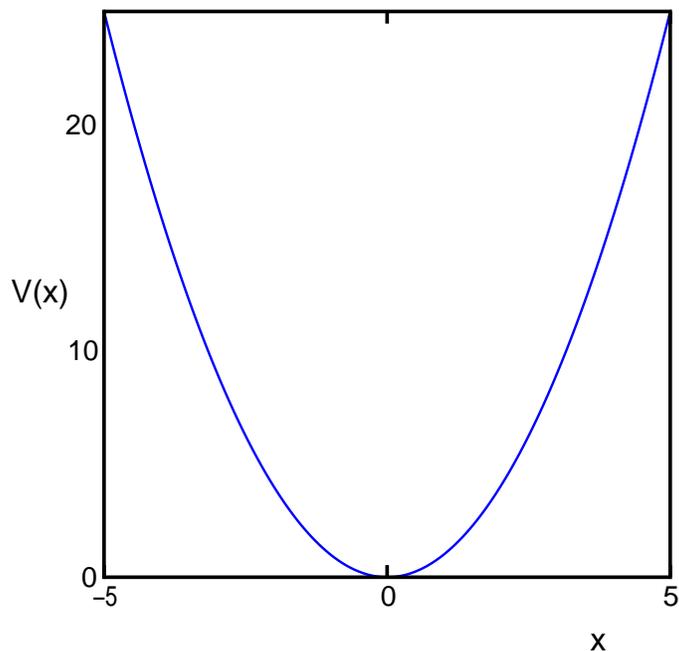} 
    \caption{The harmonic (parabolic) oscillator potential.}
    \label{fig:armonic}
\end{figure}

Since we consider a time-independent potential, the eigenfunctions
$\psi_n$ and the eigenvalues $E_n$ are obtained by means of the
time-independent Schr\"odinger equation
\begin{equation}\label{ho5}
\hat{H}\psi=E\psi~.
\end{equation}

For the HO Hamiltonian, the Schr\"odinger equation is
\begin{equation}\label{ho6}
\frac{d^2\psi}{dx^2}+\Bigg[\frac{2mE}{\hbar^2}
-\frac{m^2\omega^2}{\hbar^2}x^2\Bigg]\psi=0~.
\end{equation}
Defining the parameters
\begin{equation}\label{ho7}
k^2=\frac{2mE}{\hbar^2}~, \qquad \lambda=\frac{m\omega}{\hbar}~,
\end{equation}

\noindent the Schr\"odinger equation becomes
\begin{equation}\label{ho8}
\frac{d^2\Psi}{dx^2}+[k^2-\lambda^2x^2]\Psi=0~,
\end{equation}

\noindent which is known as Weber's differential equation in
mathematics. To solve this equation one makes use of the following
variable transformation
\begin{equation}\label{ho9}
y=\lambda x^2~.
\end{equation}

By changing the independent variable from $x$ to $y$, the
differential operators $D_x$ and $D_x^2$ take the form
\begin{equation}\label{ho10}
D_x\equiv\frac{d}{dx}=\left(\frac{dy}{dx}\right)D_y 
~, \quad D_x^2\equiv
\frac{d^2}{dx^2}=\frac{d}{dx}\left(\frac{dy}{dx}\frac{d}{dy}\right)
=\left(\frac{d^2y}{dx^2}\right)D_y+\left(\frac{dy}{dx}\right)^2D_y^2~.
\end{equation}

By applying these rules to the proposed transformation we obtain the
following differential equation in the $y$ variable
\begin{equation}\label{ho11}
y\frac{d^2\psi}{dy^2}+\frac{1}{2}\frac{d\psi}{dy}+\left[\frac{k^2}{4\lambda}
-\frac{1}{4}y\right]\psi=0~,
\end{equation}
\noindent and, by defining:
\begin{equation}\label{ho12}
\kappa=\frac{k^2}{2\lambda} 
=\frac{E}{\hbar\omega}~,
\end{equation}
we get after dividing by $y$ (i.e., $y\neq 0$)

\begin{equation}\label{ho14}
\frac{d^2\psi}{dy^2}+\frac{1}{2y}\frac{d\psi}{dy}
+\left[\frac{\kappa}{2y}-\frac{1}{4}\right]\psi=0~.
\end{equation}
Let us try to solve this equation by first doing its asymptotic
analysis in the limit $y\rightarrow\infty$ in which the equation
behaves as follows
\begin{equation}\label{ho15}
\frac{d^2\psi_{\infty}}{dy^2}-\frac{1}{4}\psi_{\infty}=0~.
\end{equation}
This equation has as solution
\begin{equation}\label{ho16}
\psi_{\infty}(y)=A\exp\left(\frac{y}{2}\right)+B\exp\left(-\frac{y}{2}\right)~.
\end{equation}

The first term diverges in the limit $y\rightarrow\infty$. Thus we
take $A=0$, and keep only the attenuated exponential. We can now
suggest that $\psi$ has the following form
\begin{equation}\label{ho17}
\psi(y)=\exp\left(-\frac{y}{2}\right)\varphi(y)~.
\end{equation}
Plugging it in the differential equation for $y$ ( Eq.~(\ref{ho14}))
one gets:
\begin{equation}\label{ho18}
y\frac{d^2\varphi}{dy^2}
+\left(\frac{1}{2}-y\right)\frac{d\varphi}{dy}+\left(\frac{\kappa}{2}-\frac{1}{4}\right)\varphi=0~.
\end{equation}
The latter equation is of the confluent (Kummer) hypergeometric form
\begin{equation}\label{ho19}
z\frac{d^2f}{dz^2}+(c-z)\frac{df}{dz}-af=0~,
\end{equation}
whose general solution is
\begin{equation}\label{ho20}
f(z)=A \hspace{.2cm} _1F_1(a,c;z)+ B \hspace{.2cm} z^{1-c}
\hspace{.1cm} _1F_1(a-c+1,2-c;z)~,
\end{equation}
where the confluent hypergeometric function is defined as follows
\begin{equation}\label{ho21}
_1F_1(a,c;z)=\sum_{n=0}^{\infty}\frac{(a)_n z^n}{(c)_n n!}~.
\end{equation}
By direct comparison of Eqs.~(\ref{ho19}) and (\ref{ho18}), one can
see that the general solution of the latter one is
\begin{equation}\label{ho22}
\varphi(y)=A\hspace{.2cm} _1F_1\left(a,\frac{1}{2};y\right)+ B
\hspace{.2cm} y^{\frac{1}{2}} \hspace{.2cm}
_1F_1\left(a+\frac{1}{2},\frac{3}{2};y\right)~,
\end{equation}
where
\begin{equation}\label{ho23}
a=-\left(\frac{\kappa}{2}-\frac{1}{4}\right)~.
\end{equation}
If we keep these solutions in their present form, the normalization
condition is not satisfied by the wavefunction. Indeed, because the
$|y|\rightarrow \infty$ asymptotic behaviour of the confluent
hypergeometric function is $_1F_1(a,c;y)\rightarrow
\frac{\Gamma(c)}{\Gamma(c-a)}e^{-ia\pi}y^{-a}
+\frac{\Gamma(c)}{\Gamma(a)}e^{y}y^{a-c}$, it follows from the
dominant exponential behavior that:
\begin{equation}\label{ho24}
\psi(y)=e^{(-\frac{y}{2})}\varphi(y)\rightarrow \hspace{.3cm}{\rm
const.} \hspace{.2cm} e^{(\frac{y}{2})}y^{a-\frac{1}{2}}~.
\end{equation}
This leads to a divergence in the normalization integral, which
physically is not acceptable. What one does in this case is to
impose the termination condition for the series. That is, the series
is cut to a finite number of $n$ terms and therefore turns into a
polynomial of order $n$. The truncation condition of the confluent
hypergeometric series $_1F_1(a,c;z)$ is $a=-n$, where $n$ is a
nonnegative integer (i.e., zero is included).
\\
We thus notice that asking for a finite normalization constant, (as
already known, a necessary condition for the physical interpretation
in terms of probabilities), leads us to the truncation of the
series, which simultaneously generates the
quantization of energy.\\
In the following we consider the two possible cases:

$1)\hspace{.4cm} a=-n \hspace{.3cm}$ and $ B=0$
\begin{equation}\label{ho25}
 \frac{\kappa}{2}-\frac{1}{4}=n~.
\end{equation}
The eigenfunctions are given by
\begin{equation}\label{ho26}
\psi_n(x)={\cal N}_{n} \exp\left(\frac{-\lambda x^2}{2}\right)
\hspace{.1cm} _1F_1\left(-n,\frac{1}{2};\lambda x^2\right)
\end{equation}

and the energy is:
\begin{equation}\label{ho27}
E_n=\hbar\omega\left(2n+\frac{1}{2}\right)~.
\end{equation}

$2)\hspace{.4cm} a+\frac{1}{2}=-n \hspace{.3cm}$ and $A=0$
\begin{equation}\label{ho28}
\frac{\kappa}{2}-\frac{1}{4}=n+\frac{1}{2}~.
\end{equation}
The eigenfunctions are now
\begin{equation}\label{ho29}
\psi_n(x)={\cal N}_{n}\exp\left(-\frac{\lambda x^2}{2}\right)
\hspace{.2cm}x \hspace{.2cm}_1F_1\left(-n,\frac{3}{2};\lambda
x^2\right)~,
\end{equation}

whereas the stationary energies are
\begin{equation}\label{ho30}
E_n=\hbar\omega\left[(2n+1)+\frac{1}{2}\right]~.
\end{equation}
The polynomials obtained by this truncation of the confluent
hypergeometric series are called Hermite polynomials and in
hypergeometric notation they are
\begin{eqnarray}\label{ho31}
H_{2n}(\eta)&=&(-1)^n \frac{(2n)!}{n!} \hspace{.2cm}
_1F_1\left(-n,\frac{1}{2};\eta^2\right) \ , \\
H_{2n-1}(\eta)&=&(-1)^n \frac{2(2n+1)!}{n!} \hspace{.2cm}\eta
\hspace{.2cm} _1F_1\left(-n,\frac{3}{2};\eta^2\right)~.
\end{eqnarray}

We can now combine the obtained results (because some of them give
us the even cases and the others the odd ones) in a single
expression for the oscillator eigenvalues and eigenfunctions
\begin{eqnarray}\label{ho33}
\psi_n (x)&=&{\cal N}_{n} \exp\left( -\frac{\lambda x^2}{2}\right)
H_n (\sqrt{\lambda}x)\\
E_n &=&\left(n+\frac{1}{2}\right)\hbar\omega
\hspace{1cm}n=0,1,2~\ldots
\end{eqnarray}

The HO energy spectrum is equidistant, i.e., there is the same
energy difference $\hbar \omega$ between any consecutive neighbor
levels. Another remark refers to the minimum value of the energy of
the oscillator; somewhat surprisingly it is not zero. This is
considered by many people to be a pure quantum result because it is
zero when $\hbar\rightarrow 0$. $E_0=\frac{1}{2}\hbar \omega$ is
known as the zero point energy and the fact that it is
nonzero is the main characteristic of all confining potentials.\\

\subsection{The Normalization Constant}

The normalization constant is usually calculated in the following
way. The Hermite generating function $e^{\lambda (-t^2+2tx)}$ is
multiplied by itself and then by $e^{-\lambda x^2}$:

\begin{equation}\label{normH1}
e^{-\lambda x^2}e^{\lambda(-s^2+2sx)}e^{\lambda(-t^2+2tx)}=\sum
_{m,n=0}^{\infty}e^{-\lambda
x^2}H_m(\sqrt{\lambda}x)H_n(\sqrt{\lambda}x)\frac{\lambda
^{\frac{m+n}{2}}s^mt^n}{m!n!}~.
\end{equation}
Integrating over $x'=\sqrt{\lambda}x$ on the whole real line, the
cross terms of the double sum drop out because of the orthogonality
property
\begin{eqnarray}\label{normH2}
\sum _{n=0}^{\infty}\frac{\lambda ^n(st)^n}{(n!)^2}\int
_{-\infty}^{\infty}e^{-x'^2}[H_n(x')]^2dx'&=&\int
_{-\infty}^{\infty}e^{-x'^2-s'^2+2s'x'-t'^2+2t'x'}dx'=\nonumber\\
& & \\
=\int _{-\infty}^{\infty}e^{-(x'-s'-t')^2}e^{2s't'}dx'&=&\pi
^{1/2}e^{2st}=\pi ^{1/2}\sum _{n=0}^{\infty}\frac{2^n\lambda
^n(st)^n}{n!}\nonumber
\end{eqnarray}
where the properties of the Euler gamma function have been used
\begin{equation}\label{defGamma}
\Gamma (z)=2\int _0^{\infty}e^{-t^2}t^{2z-1}dt~, \qquad {\rm Re}(z)
>0
\end{equation}
as well as the particular case $z=1/2$ when $\Gamma
(1/2)=\sqrt{\pi}$.

By equating coefficients of like powers of $st$ in (\ref{normH2}),
we obtain
\begin{equation}\label{normHfinal}
\int _{-\infty}^{\infty}e^{-x'^2}[H_n(x')]^2dx'=2^n\pi ^{1/2}n!~.
\end{equation}

This leads to
\begin{equation}\label{ho35}
{\cal N}_{n} = \Bigg[ \sqrt{\frac{\lambda}{\pi}}\frac{1}{2^n
n!}\Bigg]^{\frac{1}{2}}~.
\end{equation}

\medskip

\subsection{Final Formulas for the HO Stationary States}

Thus, one gets the following normalized eigenfunctions (stationary
states) of the one-dimensional harmonic oscillator operator
\begin{equation} \label{ho36}
\psi_n (x)= \Bigg[ \sqrt{\frac{\lambda}{\pi}}\frac{1}{2^n
n!}\Bigg]^{\frac{1}{2}} \hspace{.2cm} \exp \left(\frac{-\lambda
x^2}{2}\right) \hspace{.2cm} H_n( \sqrt{\lambda} x)~.
\end{equation}
If the dynamical phase factor ${\cal F}$ is included, the harmonic
oscillator eigenfunctions takes the following final form
\begin{equation} \label{ho37}
\psi_n (x,t)= \Bigg[ \sqrt{\frac{\lambda}{\pi}}\frac{1}{2^n
n!}\Bigg]^{\frac{1}{2}} \hspace{.2cm} \exp
\left(-i\left(n+\frac{1}{2}\right)\omega t-\frac{\lambda
x^2}{2}\right) \hspace{.2cm} H_n( \sqrt{\lambda} x)~.
\end{equation}

\subsection{The Algebraic Approach: Creation and Annihilation Operators
$\hat{a}^{\dagger}$ and $\hat{a}$}

There is another approach to deal with the HO besides the
conventional one of solving the Schr\"odinger equation. It is the
algebraic method, also known as the method of creation and
annihilation (ladder) operators. This is a very efficient procedure,
which can be successfully applied to many
quantum-mechanical problems, especially when dealing with discrete spectra.\\
Let us define two nonhermitic operators $a$ and $a^{\dd}$ :
\begin{equation}\label{ho37}
a=\sqrt{\frac{m\omega}{2\hbar}}\left(x+\frac{ip}{m\omega}\right) \ ,
\end{equation}
\begin{equation}\label{ho38}
a^{\dd}=\sqrt{\frac{m\omega}{2\hbar}}\left(x-\frac{ip}{m\omega}\right)~.
\end{equation}

These operators are known as 
annihilation operator and creation operator, respectively (the
reason of this terminology will be seen in the following.\\
Let us calculate the commutator of these operators
\begin{equation}\label{ho39}
[a,a^{\dd}]=\frac{m\omega}{2\hbar}\left[x
+\frac{ip}{m\omega},x-\frac{ip}{m\omega}\right]=\frac{1}{2\hbar}(-i[x,p]+i[p,x])=1~,
\end{equation}
where we have used the commutator $[x,p]=i\hbar$. Therefore the
annihilation and creation  operators do not commute, since we have
$[a,a^{\dd}]=1$. Let us also introduce the very important number
operator $\hat{N}$:
\begin{equation}\label{ho42}
\hat{N}=\ad a~.
\end{equation}
This operator is hermitic as one can readily prove using
$(AB)^{\dd}=B^{\dd}A^{\dd}$:
\begin{equation}\label{ho43}
\hat{N}^{\dd}=(\ad a)^{\dd}=\ad (\ad)^{\dd}=\ad a=\hat{N}~.
\end{equation}

Considering now that
\begin{equation}\label{ho44}
\ad a
=\frac{m\omega}{2\hbar}\left(x^2+\frac{p^2}{m^2\omega^2}\right)+\frac{i}{2\hbar}[x,p]=\frac{\hat{H}}{\hbar\omega}-\frac{1}{2}
\ ,
\end{equation}
\noindent we notice that the Hamiltonian can be written in a quite
simple form as a linear function of the number operator
\begin{equation}\label{ho45}
\hat{H}=\hbar\omega\left(\hat{N}+\frac{1}{2}\right)~.
\end{equation}
The number operator bears this name because its eigenvalues are
precisely the subindexes of the eigenfunctions on which it acts
\begin{equation}\label{ho46}
\hat{N}|n\rangle=n|n\rangle~,
\end{equation}
\noindent where we have used Dirac's ket notation $\psi _n =
|n\rangle$.

Applying the number-form of the HO Hamiltonian in (\ref{ho45}) to
this ket, one gets
\begin{equation}\label{ho48}
\hat{H}|n\rangle=\hbar\omega\left(n+\frac{1}{2}\right)|n\rangle~,
\end{equation}
which directly shows that the energy eigenvalues are given by
\begin{equation}\label{ho49}
E_n=\hbar\omega\left(n+\frac{1}{2}\right)~.
\end{equation}
Thus, this basic result is obtained through purely algebraic means.\\

It is possible to consider the ket $\ad |n\rangle$ as an eigenket of
that number operator for which the eigenvalue is raised by one unit.
In physical terms, this means that an energy quanta has been
produced by the action of $\ad$ on the ket $|n\rangle$. This already
explains the name of creation operator. Similar comments with
corresponding conclusion can be inferred for the operator $a$
explaining the name of annihilation operator (an energy quanta is
eliminated from the system when this operator is put into action).\\

Consequently, we have
\begin{equation}\label{ho62}
\ad |n\rangle=\sqrt{n+1}| n+1\rangle~.
\end{equation}

For the annihilation operator, following the same procedure one can
get the following relation
\begin{equation}\label{ho63}
a|n\rangle=\sqrt{n}| n-1\rangle~.
\end{equation}

Let us show now that the values of $n$ should be nonnegative
integers. For this, we employ the positivity requirement for the
norm of the state vector $a| n\rangle$. The latter condition tells
us that the inner product of the vector with its adjoint $
(a|n\rangle)^\dd$ (= $\langle n| \ad$) should always be nonnegative
\begin{equation}\label{ho64}
( \langle n| \ad)\cdot(a| n\rangle)\geq 0~.
\end{equation}

This relationship is in fact the expectation value of the number
operator
\begin{equation}\label{ho65}
\langle n| \ad a| n\rangle=\langle n|\hat{N}| n\rangle=\langle n|\ad
\sqrt{n}|n-1\rangle= \langle n|\sqrt{n}\sqrt{n}|n\rangle=n \geq 0~.
\end{equation}
Thus, $n$ cannot be negative. It should be a positive integer since,
if that would not be the case, by applying iteratively the
annihilation operator a sufficient number of times we would be led
to imaginary and negative
eigenvalues, which would be a contradiction to the positivity of the inner product.\\
It is possible to express the state $|n \rangle$ directly as a
function of the ground state $| 0\rangle$ using the $n$th power of
the creation operator as follows:

\begin{eqnarray}
 |n\rangle=\left[ \frac{ (\ad)^n}{\sqrt{n!}}\right]| 0\rangle ~,
\end{eqnarray}
which can be obtained by iterations.

One can also apply this method to get the eigenfunctions in the
configuration space. To achieve this, we start with the ground state
\begin{equation}\label{ho66}
a|0\rangle=0~.
\end{equation}
In the $x$-representation, we have
\begin{equation}\label{ho67}
\hat{ a} \psi_0(x)=\sqrt{\frac{m\omega}{2\hbar}}
\left(x+\frac{ip}{m\omega}\right) \psi_0(x)=0~.
\end{equation}
Recalling the form of the momentum operator in this representation,
we can obtain a differential equation for the wavefunction of the
ground state. Moreover, introducing the oscillator length
$x_0=\sqrt{\frac{\hbar}{m\omega}}=\frac{1}{\sqrt{\lambda}}$, we get
\begin{equation}\label{ho68}
\left(x+x_0^2\frac{d}{dx}\right)\psi_0=0~.
\end{equation}
This equation can be readily solved and the normalization to unity
of the full line integral of the squared modulus of the solution
leads to the physical wavefunction of the HO ground state
\begin{equation}\label{ho69}
\psi_0(x)=\left(\frac{1}{\sqrt{ \sqrt{\pi}x_0}}\right)e^{
-\frac{1}{2}(\frac{x}{x_0})^2}~.
\end{equation}
The rest of the eigenfunctions describing the HO excited states, can
be obtained by employing iteratively the creation operator. The
procedure is the following
\begin{eqnarray}
\psi_1&=&\ad \psi_0 =\left(\frac{1}{\sqrt{2}x_0}\right)\left(x-x_0^2\frac{d}{dx}\right)\psi_0 \ , \\
\psi_2&=&\frac{1}{\sqrt{2}}(\ad)^2\psi_0=\frac{1}{\sqrt{2!}}\left(\frac{1}
{\sqrt{2}x_0}\right)^2\left(x-x_0^2\frac{d}{dx}\right)^2\psi_0~.
\end{eqnarray}
By mathematical induction, one can show that
\begin{equation}\label{ho70}
\psi_n(x)=\frac{1}{\sqrt{ \sqrt{\pi}2^nn!}}\hspace{.2cm}
\frac{1}{x_0^{n+\frac{1}{2}}}
\hspace{.2cm}\left(x-x_0^2\frac{d}{dx}\right)^n
\hspace{.2cm}e^{-\frac{1}{2}(\frac{x}{x_0})^2}~.
\end{equation}

\subsection{HO Spectrum Obtained from Wilson-Sommerfeld Quantization Condition}

In the classical phase space, the equation $H=E$ for the harmonic
oscillator when divided by $E$, i.e.,
\begin{equation}\label{ws1}
\frac{p_x^2}{2mE}+\frac{m\omega _0^2x^2}{2E}=1 \ ,
\end{equation}
turns into the equation for an ellipse
\begin{equation}\label{ws2}
\frac{x^2}{a^2}+\frac{p_x^2}{b^2}=1 \ ,
\end{equation}
where $a=\frac{1}{\omega _0}\sqrt{\frac{2E}{m}}$ and $b=\sqrt{2mE}$.
Therefore, applying the Bohr-Sommerfeld quantization rule for this
case
\begin{equation}\label{ws3}
J=\oint p_xdx=\pi a b =\frac{2\pi E}{\omega _0}=2\pi \hbar n~,
\end{equation}
one obtains immediately the spectrum
\begin{equation}\label{ws4}
E_n=n\hbar\omega _0 \ ,
\end{equation}
which is the quantum HO spectrum up to the zero-point energy.

\bigskip

\section{The 3D Coulomb Well: The Stationary States of the Hydrogen Atom}

The case of the Hydrogen atom corresponds in wave mechanics to an
effective potential well that is the sum of the Coulomb well and the
quantum centrifugal barrier as shown in Fig.~(\ref{fig:PCoul}). This
result comes out from the technique of the separation of variables
that is to be considered for any differential equation in more than
one dimension. A very good introduction to this technique can be
found in the textbook of Arfken and Weber \cite{aw6}. In general,
for $d$ variables there are $d-1$ separation constants.

In spherical coordinates, the Schr\"odinger equation $(\nabla
^2_r+V(r))\psi({\bf r})=E\psi ({\bf r})$ reads
\begin{equation} 
\frac{1}{r^{2}} \frac{\partial}{\partial r}\left(r^{2}
\frac{\partial \psi}{\partial r} \right) + \frac{1}{r^{2}
\sin\theta} \frac{\partial}{\partial \theta} \left(\sin\theta
\frac{\partial \psi}{\partial \theta}\right) +
\frac{1}{r^{2}\sin^{2}\theta} \frac{\partial^{2} \psi}{\partial
\phi^{2}} + \frac{2m}{\hbar^{2}}(E - u)\psi = 0 \ ,
\end{equation}

that can be also written in the form
\begin{equation} 
\sin^{2}\theta \frac{\partial}{\partial r}\left(r^{2} \frac{\partial
\psi}{\partial r}\right) + \sin\theta \frac{\partial}{\partial
\theta}\left(\sin\theta \frac{\partial \psi}{\partial \theta}\right)
+ \frac{\partial^{2} \psi}{\partial \phi^{2}} +
\frac{2mr^{2}\sin^{2}\theta}{\hbar^{2}} \left(\frac{e^{2}}{4\pi
\epsilon_{0}r} + E\right)\psi = 0~.
\end{equation}
This equation is a partial differential equation for the electron
wavefunction $\psi(r,\theta,\phi)$ `within' the atomic hydrogen.
Together with the various conditions that the wavefunction
$\psi(r,\theta,\phi)$ should fulfill [for example,
$\psi(r,\theta,\phi)$ should have a unique value at any spatial
point ($r,\theta,\phi$)], this equation specifies in a complete
manner the stationary behavior of the hydrogen electron.\\

\begin{figure}
\centering
    \includegraphics[width=4.3in]{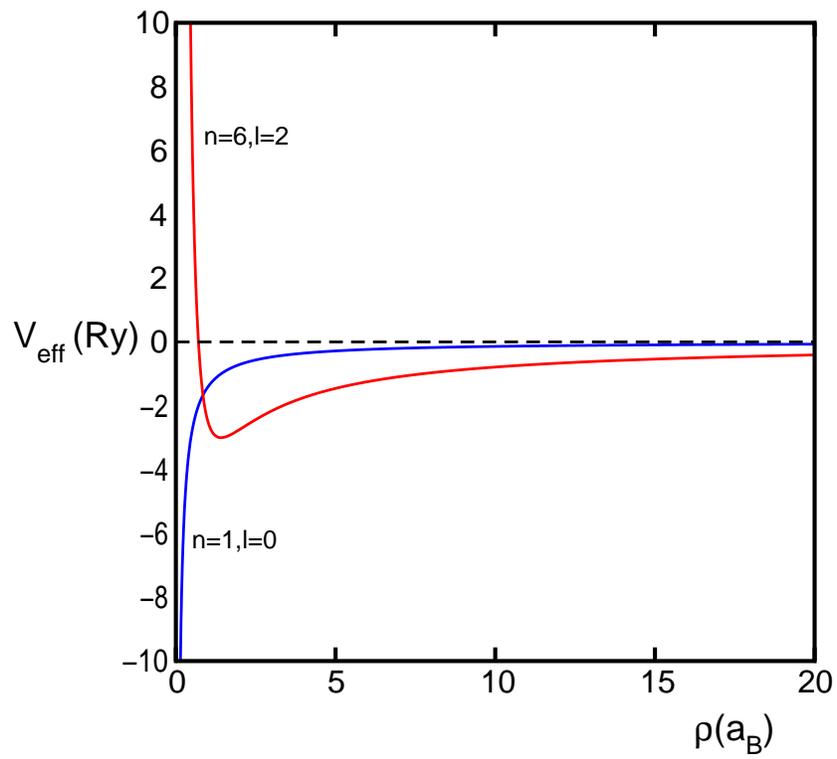} 
    \caption{The effective potential well in the case of the hydrogen atom consisting of the
electrostatic potential plus a quantized centrifugal barrier (see
text).}
    \label{fig:PCoul}
\end{figure}

\subsection{The Separation of Variables in Spherical Coordinates}

The real usefulness of writing the hydrogen Schr\"odinger equation
in spherical coordinates consists in the easy way of achieving the
separation procedure in three independent one-dimensional equations.
The separation procedure is to seek the solutions for which the
wavefunction $\psi(r,\theta,\phi)$ has the form of a product of
three functions, each of one of the three spherical variables,
namely $R(r)$ depending only on $r$, $\Theta(\theta)$ depending only
on $\theta$, and $\Phi(\phi)$ that depends only on $\phi$. This is
quite similar to the separation of the Laplace equation. Thus

\begin{equation} 
\psi(r,\theta,\phi) = R(r)\Theta(\theta)\Phi(\phi)~.
\end{equation}

The $R(r)$ function describes the differential variation of the
electron wavefunction $\psi$ along the vector radius coming out from
the nucleus, with $\theta$ and $\phi$ assumed to be constant. The
differential variation of $\psi$ with the polar angle $\theta$ along
a meridian of an arbitrary sphere centered in the nucleus is
described only by the function $\Theta(\theta)$ for constant $r$ and
$\phi$. Finally, the function $\Phi(\phi)$ describes how $\psi$
varies with the azimuthal angle $\phi$ along a parallel of an
arbitrary sphere centered at the nucleus, under the conditions that
$r$ and $\theta$ are kept constant.

Using $\psi=R\Theta\Phi$, one can see that

\begin{equation} 
\frac{\partial \psi}{\partial r} = \Theta \Phi \frac{d R}{d
r}~,\qquad \frac{\partial \psi}{\partial \theta} = R\Phi \frac{d
\Theta}{d \theta}~,\qquad \frac{\partial \psi}{\partial \phi} =
R\Theta \frac{d \Phi}{d\phi }~.
\end{equation}

Then, one can obtain the following equations for the three factoring
functions:
\begin{equation} \label{91}
\frac{d^{2}\Phi}{d\phi^{2}} + m_{l}^{2}\Phi = 0~,
\end{equation}
\begin{equation} \label{92}
\frac{1}{\sin\theta}\frac{d}{d\theta}\left(\sin\theta
\frac{d\Theta}{d\theta}\right) +
\left[l(l+1)-\frac{m_{l}^{2}}{\sin^{2}\theta}\right]\Theta = 0~,
\end{equation}
\begin{equation} \label{93} 
\frac{1}{r^{2}}\frac{d}{dr}\left(r^{2}\frac{dR}{dr}\right) +
\left[\frac{2m}{\hbar^{2}}\left(\frac{e^{2}}{4\pi \epsilon_{0}r} +
E\right) - \frac{l(l+1)}{r^{2}}\right]R = 0~.
\end{equation}

Each of these equations is an ordinary differential equation for a
function of a single variable. In this way, the Schr\"odinger
equation for the hydrogen electron, which initially was a partial
differential equation for a function $\psi$ of three variables, gets
a simple form of three 1D ordinary differential equations for
unknown functions of one variable. The reason why the separation
constants have been chosen as $-m_l^2$ and $-l(l+1)$ will become
clear in the following subsections.

\subsection{The Angular Separation Constants as Quantum Numbers}

\subsubsection{The Azimuthal Solution and the Magnetic Quantum Number}

The Eq.~(\ref{91}) is easily solved leading to the following
solution
\begin{equation} 
\Phi(\phi) = {\cal N}_{\Phi}e^{im_{l}\phi}~,
\end{equation}
where ${\cal N}_{\Phi}$ is the integration constant that will be
used as a normalization constant for the azimuthal part. One of the
conditions that any wavefunctions should fulfill is to have a unique
value for any point in space. This applies to $\Phi$ as a component
of the full wavefunction $\psi$. One should notice that\ $\phi$ and
$\phi + 2\pi$ must be identical in the same meridional plane.
Therefore, one should have $\Phi(\phi)= \Phi(\phi + 2\pi)$, i.e.,
${\cal N}_{\Phi}e^{im_{l}\phi} = {\cal N}_{\Phi}e^{im_{l}(\phi +
2\pi)}$. This can be fulfilled only if $m_{l}$ is zero or a positive
or negative integer $(\pm 1, \pm 2, \pm 3,...)$. The number $m_{l}$
is known as the magnetic quantum number of the atomic electron and
is related to the direction of the projection of the orbital
momentum $L_{z}$. It comes into play whenever the effects of axial
magnetic fields on the electron may show up. There is also a deep
connection between $m_{l}$ and the orbital quantum number $l$, which
in turn determines the modulus of the orbital momentum of the
electron.

The solution for $\Phi$ should also fulfill the normalization
condition when integrating over a full period of the azimuthal
angle,

\begin{equation} 
\int_{0}^{2\pi} \mid \Phi \mid^{2}d\phi = 1
\end{equation}
and substituting $\Phi$, one gets
\begin{equation} 
\int_{0}^{2\pi} {\cal N}_{\Phi}^{2}d\phi = 1~.
\end{equation}
It follows that ${\cal N}_{\Phi}=1/\sqrt{2\pi}$, and therefore the
normalized $\Phi$ is
\begin{equation} 
\Phi(\phi) = \frac{1}{\sqrt{2\pi}}e^{im_{l}\phi}~.
\end{equation}

\subsubsection{The Polar Solution and the Orbital Quantum Number}

The solution of the $\Theta(\theta)$ equation is more complicated
since it contains two separation constants which can be proved to be
integer numbers. Things get easier if one reminds that the same
Eq.~(\ref{92}) occurs also when the Helmholtz equation for the
spatial amplitude profiles of the electromagnetic normal modes is
separated in spherical coordinates. From this case we actually know
that this equation is the associated Legendre equation for which the
polynomial solutions are the associated Legendre polynomials
\begin{equation} \label{theta1}
P_{l}^{m_{l}}(x) = (-1)^{m_{l}}(1-x^{2})^{m_{l}/2}
\frac{d^{m_{l}}}{dx^{m_{l}}}P_{l}(x) =
(-1)^{m_{l}}\frac{(1-x^{2})^{m_{l}/2}}{2^{l}l!}\frac{d^{m_{l} +
l}}{dx^{{m_{l} + l}}}(x^{2} - 1)^{l}~.
\end{equation}
The function $\Theta(\theta)$ are a normalized form,
$\Theta(\theta)={\cal N}_{\Theta}P_{l}^{m_l}$, of the associated
Legendre polynomials
\begin{equation} \label{theta2} 
\Theta(\theta) = \sqrt{\frac{2l+1}{2}\frac{(l-m_{l})!}{(l+m_{l})!}}
P_{l}^{m_{l}}(cos\theta)~.
\end{equation}

For the purposes here, the most important property of these
functions is that they exist only when the constant $l$ is an
integer number greater or at least equal to $|m_{l}|$, which is the
absolute value of $m_{l}$. This condition can be written in the form
of the set of values available for $m_{l}$
\begin{equation} \label{theta3}
m_{l} = 0,\pm 1, \pm 2,...,\pm l~.
\end{equation}
The condition that $l$ should be a positive integer can be seen from
the fact that for noninteger values, the solution of Eq.~(\ref{92})
diverges for $\cos \theta =\pm 1$, while for physical reasons we
require finite solutions in these limits. The other condition $l\geq
|m_l|$ can be obtained from examining Eq.~({\ref{theta1}), where one
can see a derivative of order $m_l+l$ applied to a polynomial of
order $2l$. Thus $m_l+l$ cannot be greater than $2l$. On the other
hand, the derivative of negative order are not defined and that of
zero order is interpreted as the unit operator. This leads to
$m_l\geq -l$.

The interpretation of the orbital number $l$ does have some
difficulties. Let us examine the equation corresponding to the
radial wavefunction $R(r)$. This equation rules only the radial
motion of the electron, i.e., with the relative distance with
respect to the nucleus along some guiding ellipses. However, the
total energy of the electron $E$ is also present. This energy
includes the kinetic electron energy in its orbital motion that is
not related to the radial motion. This contradiction can be
eliminated using the following argument. The kinetic energy $T$ has
two parts: a pure radial one $T_{radial}$ and $T_{orbital}$, which
is due to the closed orbital motion. The potential energy $V$ of the
electron is the attractive electrostatic energy. Therefore, the
electron total energy is
\begin{equation}  
E = T_{radial} + T_{orbital} - \frac{e^{2}}{4\pi \epsilon_{0}r}~.
\end{equation}
Substituting this expression of $E$ in Eq.~(\ref{93}) we get after
some regrouping of the terms

\begin{equation}  
\frac{1}{r^{2}}\frac{d}{dr}\left(r^{2}\frac{dR}{dr}\right) +
\frac{2m}{\hbar^{2}}\left[T_{radial} + T_{orbital} -
\frac{\hbar^{2}l(l+1)}{2mr^{2}}\right]R=0~.
\end{equation}

If the last two terms in parentheses compensate each other, we get a
differential equation for the pure radial motion. Thus, we impose
the condition
\begin{equation}  
T_{orbital} = \frac{\hbar^{2}l(l+1)}{2mr^{2}}~.
\end{equation}

However, the orbital kinetic energy of the electron is $T_{orbital}
= \frac{1}{2}mv_{orbital}^{2}$ and since the orbital momentum of the
electron is $L = mv_{orbital}r$, we can express the orbital kinetic
energy in the form

\begin{equation}  
T_{orbital} = \frac{L^{2}}{2mr^{2}}~.
\end{equation}

Therefore, we have
\begin{equation}  
\frac{L^{2}}{2mr^{2}} = \frac{\hbar^{2}l(l+1)}{2mr^{2}} \ ,
\end{equation}

and consequently

\begin{equation}\label{42}  
L = \sqrt{l(l+1)}\hbar~.
\end{equation}

The interpretation of this result is that since the orbital quantum
number $l$ is constrained to take the values $l=0,1,2,...,(n-1)$,
the electron can only have orbital momenta specified by means of
Eq.~({\ref{42}). As in the case of the total energy $E$, the angular
momentum is conserved and gets quantized. Its natural unit in
quantum mechanics is $\hbar=h/2\pi=1.054 \times 10^{-34}$ J.s.

In the macroscopic planetary motion (putting aside the many-body
features), the orbital quantum number is so large that any direct
experimental detection of the quantum orbital momentum is
impossible. For example, an electron with $l= 2$ has an angular
momentum $L=2.6 \times 10^{-34}$ J.s., whereas the terrestrial
angular momentum is $2.7 \times 10^{40}$ J.s.!

A common notation for the angular momentum states is by means of the
letter $s$ for $l=0$, $p$ for $l=1$, $d$ for $l=2$, and so on. This
alphabetic code comes from the empirical spectroscopic
classification in terms of the so-called series, which was in use
before the advent of wave mechanics.

On the other hand, for the interpretation of the magnetic quantum
number, we must take into account that the orbital momentum is a
vector operator and therefore one has to specify its direction,
sense, and modulus. $L$, being a vector, is perpendicular on the
plane of rotation. The geometric rules of the vectorial products
still hold, in particular the rule of the right hand: its direction
and sense are given by the right thumb whenever the other four
fingers point at the direction of rotation.

\medskip

\subsubsection{The Space Quantization}

\medskip

We have already seen the spatial quantization of the Bohr-Sommerfeld
electron trajectories. But what significance can be associated to a
direction and sense in the limited volume of the atomic hydrogen in
Schr\"odinger wave mechanics ? The answer may be quick if we think
that the rotating electron is nothing but a one-electron loop
current that considered as a magnetic dipole has a corresponding
magnetic field. Consequently, an atomic electron will always
interact with an applied magnetic field ${\bf H}$. The magnetic
quantum number $m_{l}$ specifies the spatial direction of $L$, which
is determined by the component of $L$ along the direction of the
external magnetic field. This effect is commonly known as the
quantization of the space in a magnetic field.

If we choose the direction of the magnetic field as the $z$ axis,
the component of $L$ along this direction is
\begin{equation} 
L_{z} = m_{l}\hbar~.
\end{equation}
The possible values of $m_{l}$ for a given value of $l$, go from
$+l$ to $-l$, passing through zero, so that there are $2l+1$
possible orientations of the angular momentum $L$ in a magnetic
field. When $l=0$, $L_{z}$ can be only zero; when$l=1$, $L_{z}$ can
be $\hbar$, 0, or $-\hbar$; when $l=2$, $L_{z}$ takes only one of
the values $2\hbar$, $\hbar$, 0, $-\hbar$, or $-2\hbar$, and so
forth. It is worth mentioning that $L$ cannot be put exactly
parallel or anti-parallel to ${\bf H}$, because $L_{z}$ is always
smaller than the modulus $\sqrt{l(l+1)}\hbar$ of the total orbital
momentum.

One should consider the atom/electron characterized by a given
$m_{l}$ as having the orientation of its angular momentum $L$
determined relative to the external applied magnetic field.

In the absence of the external magnetic field, the direction of the
$z$ axis is fully arbitrary. Therefore, the component of $L$ in any
arbitrary chosen direction is $m_{l}\hbar$; the external magnetic
field offers a preferred reference direction from the experimental
viewpoint.

Why only the component $L_{z}$ is quantized ? The answer is related
to the fact that $L$ cannot be put along a direction in an arbitrary
way. There is a special precessional motion in which its `vectorial
arrow' moves always along a cone centered on the quantization axis
such that its projection $L_{z}$ is $m_{l}\hbar$. The reason why
this quantum precession occurs is different from the macroscopic
planetary motion as it is due to the uncertainty principle. If $L$
would be fixed in space, in such a way that $L_{x}$, $L_{y}$ and
$L_{z}$ would have well-defined values, the electron would have to
be confined to a well-defined plane. For example, if $L$ would be
fixed along the $z$ direction, the electron tends to maintain itself
in the plane $xy$.

This can only occur in the case in which the component $p_{z}$ of
the electron momentum is `infinitely' uncertain. This is however
impossible if the electron is part of the hydrogen atom. But since
in reality just the component $L_{z}$ of $L$ together with $L^2$
have well defined values and $|L| > |L_{z}|$, the electron is not
constrained to a single plane. If this would be the case, an
uncertainty would exist in the coordinate $z$ of the electron. The
direction of $L$ changes continuously so that the mean values of
$L_{x}$ and $L_{y}$ are zero, although $L_{z}$ keeps all the time
its value $m_{l}\hbar$. It is here where Heisenberg's uncertainty
principle helps to make a clear difference between atomic wave
motion and Bohr-Sommerfeld quantized ellipses.

\subsection{Polar and Azimuthal Solutions Set Together}

The solutions of the azimuthal and polar parts can be unified within
spherical harmonics functions that depend on both $\phi$ and
$\theta$. This simplifies the algebraic manipulations of the full
wave functions $\psi(r,\theta,\phi)$. Spherical harmonics are given
by
\begin{equation} 
Y_{l}^{m_{l}}(\theta,\phi) = (-1)^{m_{l}} \sqrt{\frac{2l+1}{4\pi}
\frac{(l-m_{l})!}{(l+m_{l})!}}
P_{l}^{m_{l}}(cos\theta)e^{im_{l}\phi}~.
\end{equation}
The factor $(-1)^{m_{l}}$ does not produce any problem because the
Schr\"odinger equation is linear and homogeneous. This factor is
added for the sake of convenience in angular momentum studies. It is
known as the Condon-Shortley phase factor and its effect is to
introduce the alternated sequence of the signs $\pm$ for the
spherical harmonics of a given $l$.

\subsection{The Radial Solution and the Principal Quantum Number}

There is no energy parameter in the angular equations and that is
why the angular motion does not make any contribution to the
hydrogen spectrum. It is the radial motion that determines the
energy eigenvalues. The solution for the radial part $R(r)$ of the
wave function $\psi$ of the hydrogen atom is somewhat more
complicated although the presence of two separation constants, $E$
and $l$, point to some associated orthogonal polynomials. In the
radial motion of the hydrogen electron significant differences with
respect to the electrostatic Laplace equation do occur. The final
result is expressed analytically in terms of the associated Laguerre
polynomials (Schr\"odinger 1926). The radial equation can be solved
exactly only when E is positive or for one of the following negative
values $E_{n}$ (in which cases, the electron is in a bound
stationary state within atomic hydrogen)
\begin{equation} 
E_{n} = 
-{\rm Ry}\left(\frac{1}{n^{2}}\right)~,
\end{equation}
where ${\rm Ry}=-\frac{m
e^{4}}{32\pi^{2}\epsilon_{0}^{2}\hbar^{2}}=13.606$ eV is the Rydberg
atomic energy unit connected with the spectroscopic Rydberg constant
$R_{\infty}$ through Ry $=hcR_{\infty}$, whereas $n$ is a positive
integer number called the principal quantum number. It gives the
quantization of the electron energy in the hydrogen atom. This
discrete atomic spectrum has been first obtained in 1913 by Bohr
using semi-empirical quantization methods and next by Pauli and
Schr\"odinger almost simultaneously in 1926.

Another condition that should be satisfied in order to solve the
radial equation is that $n$ have to be strictly bigger than $l$. Its
lowest value is $l+1$ for a given $l$. Vice versa, the condition on
$l$ is
\begin{equation} 
l = 0,1,2,...,(n-1) \ ,
\end{equation}
for given $n$.

The radial equation can be written in the form
\begin{equation} 
r^{2}\frac{d^{2}R}{dr^{2}} + 2r\frac{dR}{dr} + \left[\frac{2m
E}{\hbar^{2}}r^{2} + \frac{2me^{2}}{4\pi \epsilon_{0} \hbar^{2}}r -
l(l+1)\right]R = 0~.
\end{equation}
Dividing by $r^2$ and using the substitution $\chi (r) =rR$ to
eliminate the first derivative $\frac{dR}{dr}$, one gets the
standard form of the radial Schr\"odinger equation displaying the
effective potential $U(r)=-{\rm const}/r + l(l+1)/r^2$ (actually,
electrostatic potential plus quantized centrifugal barrier). These
are necessary mathematical steps in order to discuss a new boundary
condition, since the spectrum is obtained by means of the $R$
equation. The difference between a radial Schr\"odinger equation and
a full-line one is that a supplementary boundary condition should be
imposed at the origin ($r=0$). The Coulomb potential belongs to a
class of potentials that are called weak singular for which ${\rm
lim} _{r\rightarrow 0}\,U(r)r^2=0$. In these cases, one tries
solutions of the type $\chi \propto r^{\nu}$, implying $\nu (\nu
-1)=l(l+1)$, so that the solutions are $\nu _1 =l+1$ and $\nu
_2=-l$, just as in electrostatics. The negative solution is
eliminated for $l\neq 0$ because it leads to a divergent
normalization constant, nor did it respect the normalization at the
delta function for the continuous part of the spectrum. On the other
hand, the particular case $\nu _2 =0$ is eliminated because the mean
kinetic energy is not finite. The final conclusion is that $\chi
(0)=0$ for any $l$.

Going back to the analysis of the radial equation for $R$, the first
thing to do is to write it in nondimensional variables. This is
performed by noticing that the only space and time scales that one
can form on combining the three fundamental constants entering this
problem, namely $e^2$, $m_{e}$ and $\hbar$ are the Bohr atomic
radius $a_{B}=\hbar ^2/me^2=0.529\cdot 10 ^{-8}$ cm. and the atomic
time $t_{B}=\hbar ^3/me^4=0.242 \cdot 10^{-16}$ sec., usually known
as atomic units. Employing these units, one gets
\begin{equation} 
\frac{d^{2}R}{dr^{2}} + \frac{2}{r}\frac{dR}{dr} + \left[2 E +
\frac{2}{r} - \frac{l(l+1)}{r^2}\right]R = 0~,
\end{equation}

where we are especially interested in the discrete part of the
spectrum ($E<0$). The notations $n=1/\sqrt{-E}$ and $\rho=2r/n$
leads us to

\begin{equation} 
\frac{d^{2}R}{d\rho ^{2}} + \frac{2}{\rho}\frac{dR}{d\rho} +
\left[\frac{n}{\rho}-\frac{1}{4} - \frac{l(l+1)}{\rho ^2}\right]R =
0~.
\end{equation}

For $\rho \rightarrow \infty$, this equation reduces to
$\frac{d^{2}R}{d\rho ^{2}}=\frac{R}{4}$, having solutions $R\propto
e^{\pm\rho /2}$. Because of the normalization condition only the
decaying exponential is acceptable. On the other hand, the
asymptotic at zero, as we already commented on, should be $R\propto
\rho ^{l}$. Therefore, we can write $R$ as a product of three radial
functions $R=\rho ^{l}e^{-\rho /2}F(\rho)$, of which the first two
give the asymptotic behaviors, whereas the third is the radial
function in the intermediate region. The latter function is of most
interest because its features determine the energy spectrum. The
equation for $F$ is
\begin{equation} 
\rho\frac{d^{2}F}{d\rho ^{2}} + (2l+2-\rho)\frac{dF}{d\rho} +
(n-l-1)F = 0~.
\end{equation}
This is a particular case of confluent hypergeometric equation. It
can be identified as the equation for the associated Laguerre
polynomials $L_{n+l}^{2l+1}(\rho)$. Thus, the normalized form of $R$
is
\begin{equation} 
R_{nl}(\rho) = -\frac{2}{n^2}\sqrt{\frac{(n-l-1)!}{2n[(n+l)!]^{3}}}
e^{-\rho /2}\rho^{l} L_{n+l}^{2l+1}(\rho)~,
\end{equation}
where the following Laguerre normalization condition has been used

\begin{equation} 
\int_{0}^{\infty}e^{-\rho}\rho^{2l+1}[L_{n+l}^{2l+1}(\rho)]^{2}d\rho
= \frac{2n[(n+l)!]^{3}}{(n-l-1)!}~.
\end{equation}

\medskip

\subsection{Final Formulas for the Hydrogen Atom Stationary States}

We have now the solutions of all the equations depending on a single
variable and therefore we can build the stationary wave functions
for any electronic state of the hydrogen atom. They have the
following analytic form
\begin{equation} 
\psi(r,\theta,\phi)={\cal N}_{H}(\alpha _n r)^{l} e^{-\alpha _n r/2}
L_{n+l}^{2l+1}(\alpha _n r) P_{l}^{m_{l}}(cos\theta)e^{im_{l}\phi}~,
\end{equation}
where
\begin{equation}\label{normH}
{\cal N}_{H}=-\frac{2}{n^2}
\sqrt{\frac{2l+1}{4\pi}\frac{(l-m_{l})!}{(l+m_{l})!}
\frac{(n-l-1)!}{[(n+l)!]^{3}}}, \qquad \alpha _n=2/na_{B}~.
\end{equation}

Using the spherical harmonics, the solution is written as follows
\begin{equation}  
\psi(r,\theta,\phi)=-\frac{2}{n^2}\sqrt{\frac{(n-l-1)!}{[(n+l)!]^{3}}}
(\alpha _n r)^{l} e^{-\alpha _n r/2} L_{n+l}^{2l+1}(\alpha _n
r)Y_{l}^{m_{l}}(\theta,\phi)~.
\end{equation}

If the dynamical factor ${\cal F}$ is included, we get
\begin{equation}\label{final2}
\psi(r,\theta,\phi;t)=-\frac{2}{n^2}\sqrt{\frac{(n-l-1)!}{[(n+l)!]^{3}}}
(\alpha _n r)^{l}
e^{\left(-i\left(\frac{Ry}{\hbar}\right)\frac{t}{n^2}-
\frac{r}{na_B}\right)} L_{n+l}^{2l+1}(\alpha _n
r)Y_{l}^{m_{l}}(\theta,\phi)~.
\end{equation}

The latter formulas may be considered as the final result for the
Schr\"odinger solution of the hydrogen atom for any stationary
electron state. Indeed, one can see explicitly both the asymptotic
dependence and the two orthogonal and complete sets of functions,
i.e., the associated Laguerre polynomials and the spherical
harmonics that correspond to this particular case of linear partial
second-order differential equation. For the algebraic approach to
the hydrogen atom problem we recommend the paper of Kirchberg and
collaborators \cite{K-03} and the references therein. Finally, the
stationary hydrogen eigenfunctions are characterized by a high
degeneracy since they depend on three quantum numbers whereas the
energy spectrum is only $n$-dependent. The degree of degeneracy is
easily calculated if one notice that there are $2l+1$ values of
$m_l$ for a given $l$ which in turn takes values from $0$ to $n-1$
for a given $n$. Thus, there are $\sum _{l=0}^{n-1}(2l+1)=n^2$
states of given energy. Not only the presence of many degrees of
freedom is the cause of the strong degeneracy. In the case of the
hydrogen atom the existence of the conserved Runge-Lenz vector
(commuting operator)${\bf K}=(2me^2)^{-1}[{\bf L}\times {\bf P}-{\bf
P}\times {\bf L}]+{\bf r}/r$ introduces more symmetry into the
problem and enhances the degeneracy. On the other hand, the
one-dimensional quantum wavefunctions are not degenerate because
they are characterized by a single discrete index. The general
problem of degeneracies is nicely presented in a paper by Shea and
Aravind \cite{sa96}.

\medskip

\subsection{Electronic Probability Density}

In the Bohr model of the hydrogen atom, the electron rotates around
the nucleus on circular or elliptic trajectories. It is possible to
think of appropriate experiments allowing to ``see" that the
electron moves within experimental errors at the predicted radii
$r_n=n^{2}a_{0}$
in the equatorial plane $\theta=90^{o}$, whereas the azimuthal angle
may vary according to the specific experimental conditions.

It is in this case that the more general wave mechanics changes the
conclusions of the Bohr model in at least two important aspects:

\medskip

\noindent $\bullet$ First, one cannot speak about exact values of
$r,\theta,\phi$ (and therefore of planetary trajectories), but only
of relative probabilities to find the electron within a given
infinitesimal region of space. This feature is a consequence of the
wave nature of the electron.

\medskip

\noindent $\bullet$ Secondly, the electron does not move around the
nucleus in the classical conventional way because the probability
density $|\psi|^{2}$ does not depend on time but can vary
substantially as a function of the relative position of the
infinitesimal region.

\medskip

The hydrogen electron wave function $\psi$ is $\psi=R\Theta\Phi$,
where $R=R_{nl}(r)$ describes the way $\psi$ changes with $r$ when
the principal and orbital quantum numbers have the values $n$ and
$l$, respectively. $\Theta=\Theta_{lm_{l}}(\theta)$ describes in
turn how $\psi$ varies with $\theta$ when the orbital and magnetic
quantum numbers have the values $l$ and $m_{l}$, respectively.
Finally, $\Phi=\Phi_{m_{l}}(\phi)$ gives the change of $\psi$ with
$\phi$ when the magnetic quantum number has the value $m_{l}$. The
probability density $\mid \psi \mid^{2}$ can be written

\begin{equation}  
\mid \psi \mid^{2} = \mid R \mid^{2} \mid \Theta \mid^{2} \mid \Phi
\mid^{2}~.
\end{equation}

Notice that the probability density $|\Phi|^{2}$, which measures the
possibility to find the electron at a given azimuthal angle $\phi$,
is a constant (does not depend on $\phi$). Therefore, the electronic
probability density is symmetric with respect to the $z$ axis and
independent on the magnetic substates (at least until an external
magnetic field is applied). Consequently, the electron has an equal
probability to be found in any azimuthal direction. The radial part
$R$ of the wave function, contrary to $\Phi$, not only varies with
$r$, but it does it differently for any different combination of
quantum numbers $n$ and $l$. Figure (\ref{hydrogen3}) shows plots of
$R$ as a function of $r$ for the states $1s$, $2s$, and $2p$. $R$ is
maximum at the center of the nucleus ($r=0$) for all the $s$ states,
whereas it is zero at $r=0$ for all the states of nonzero angular
momentum.

\begin{figure}
\centering
    \includegraphics[width=4.5in]{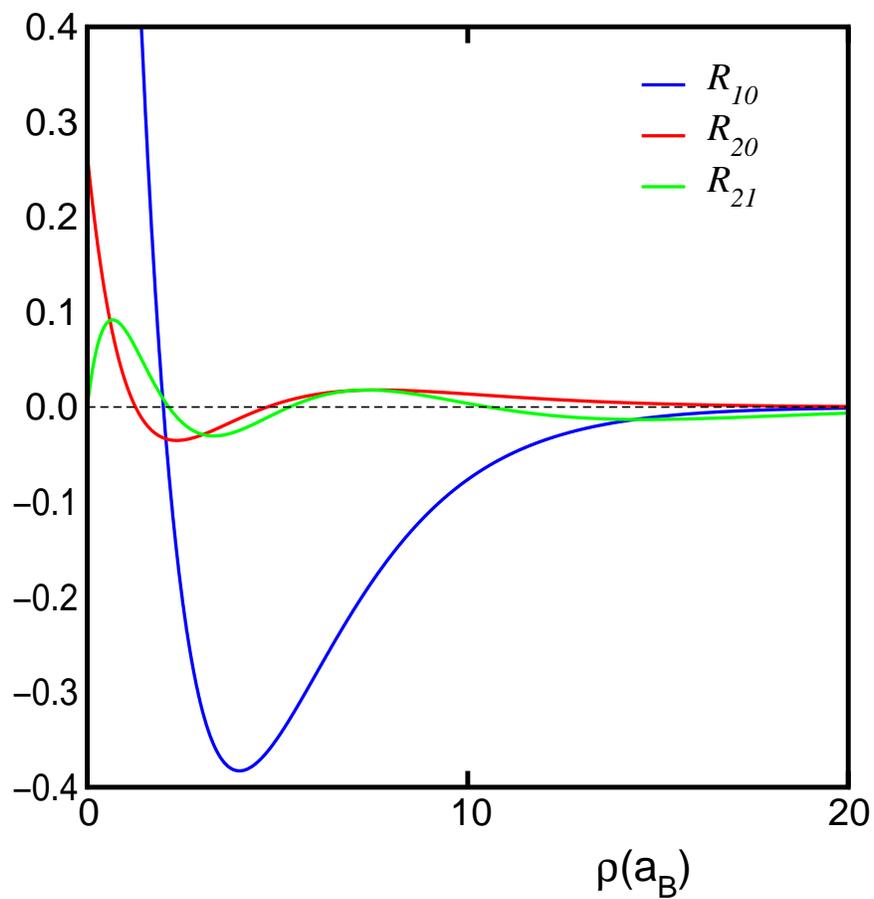}        
    \caption{Approximate plots of the radial functions
$R_{1s}$, $R_{2s}$, $R_{2p}$; ($a_B=0.53$ \AA ).}
    \label{hydrogen3}
\end{figure}

\begin{figure}
\centering
    \includegraphics[width=4.5in]{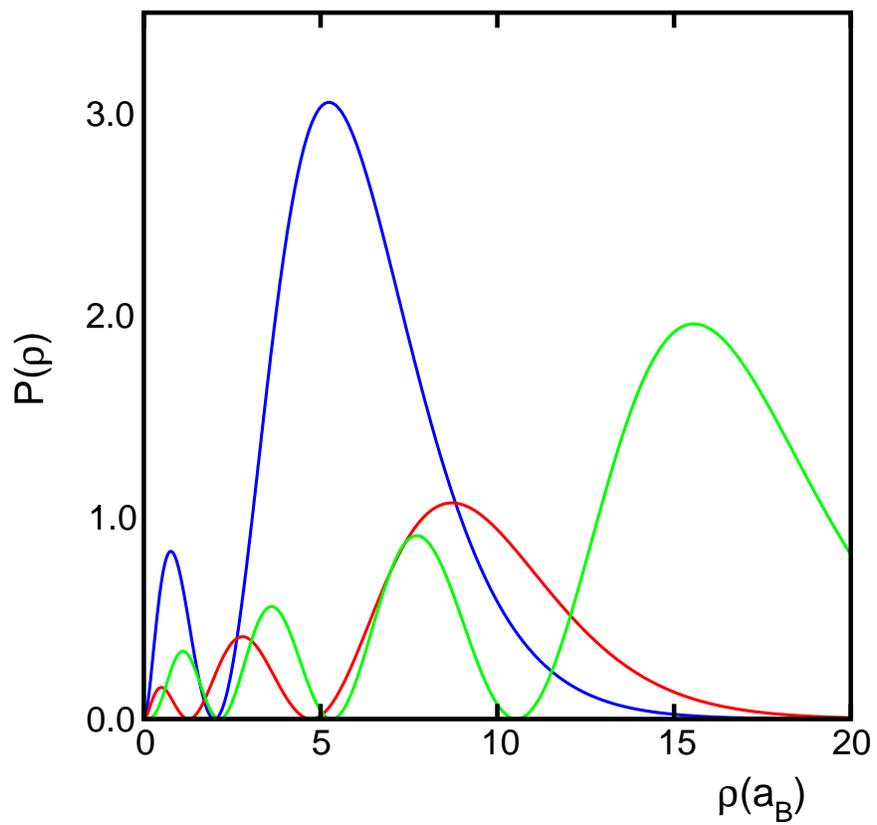}            
    \caption{Probability density of finding the hydrogen
electron between $r$ and $r+dr$ with respect to the nucleus for the
states $1s$ (blue), $2s$ (red), $2p$ (green).}
    \label{hydrogen4}
\end{figure}

The electron probability density at the point $r,\theta,\phi$ is
proportional to $|\psi|^{2}$, but the real probability in the
infinitesimal volume element $dV$ is $\mid \psi \mid^{2}dV$. In
spherical coordinates $dV=r^{2}\sin\theta dr d\theta d\phi$ and
since $\Theta$ and $\Phi$ are normalized functions, the real
numerical probability $P(r)dr$ to find the electron at a relative
distance with respect to the nucleus between $r$ and $r+dr$ is

\begin{eqnarray}  
P(r)dr & = & r^{2}\mid R \mid^{2}dr \int_{0}^{\pi} \mid\ \Theta
\mid^{2} \sin\theta d\theta \int_{0}^{2\pi}
\mid\ \Phi \mid^{2}d\phi \nonumber\\
& = & r^{2}\mid R \mid^{2}dr \ .
\end{eqnarray}

The function $P(r)$ is displayed in Fig. (\ref{hydrogen4}) for the
same states for which the radial functions $R$ is displayed in Fig.
(\ref{hydrogen3}). In principle, the curves are quite different. We
immediately see that $P(r)$ is not maximal in the nucleus for the
states $s$, as happens for $R$. Instead, their maxima are
encountered at a finite distance from the nucleus. The most probable
value of $r$ for a $1s$ electron is exactly $a_{B}$, the Bohr
radius. However, the mean value of $r$ for a $1s$ electron is
$1.5a_{B}$. At first sight this might look strange, because the
energy levels are the same both in quantum mechanics and in Bohr's
model. This apparent outmatching is eliminated if one takes into
account that the electron energy depends on $1/r$ and not on $r$,
and the mean value of $1/r$ for a $1s$ electron is exactly
$1/a_{0}$.

The function $\Theta$ varies with the polar angle $\theta$ for all
the quantum numbers $l$ and $m_{l}$, unless $l=m_{l}=0$, which are
the $s$ states. The probability density $| \Theta |^{2}$ for a $s$
state is a constant (1/2). This means that since $|\Phi|^{2}$ is
also a constant, the electronic probability density $| \psi |^{2}$
has the same value for a given value of $r$, not depending on the
direction. In other states, the electrons present an angular
behavior that in many cases may be quite complicated.
Because $|\psi |^{2}$ is independent of $\varphi$, a
three-dimensional representation of $|\psi|^{2}$ can be obtained by
rotating a particular representation around a vertical axis. This
can prove visually that the probability densities for the $s$ states
have spherical symmetry, while all the other states do not possess
it. In this way, one can get more or less pronounced lobes of
characteristic forms depending on the atomic state. These lobes are
quite important in chemistry for specifying the atomic interaction
in the molecular bulk.

\subsection{Other 3D Coordinate Systems Allowing Separation of Variables}

A complete discussion of the 3D coordinate systems allowing the
separation of variables for the Schr\"odinger equation has been
provided by Cook and Fowler \cite{cf}. Here, we briefly review these
coordinate systems:

\indent {\em Parabolic}.

\noindent The parabolic coordinates given by
\begin{eqnarray}\label{parab}
x&=&\sqrt{\xi \eta}\cos \phi~,\nonumber \\
y&=&\sqrt{\xi \eta}\sin \phi~, \\
z&=&\frac{\xi -\eta}{2}~,\nonumber
\end{eqnarray}

where $\xi \in [0,\infty )$, $\eta \in [0 , \infty )$, and $\phi \in
[0, 2\pi )$, are another coordinate system in which the
Schr\"odinger hydrogen equation is separable as first shown by
Schr\"odinger \cite{parabolic}. The final solution in this case is
expressed as the product of factors of asymptotic nature, azimuthal
harmonics, and two sets of associate Laguerre polynomials in the
variables $\xi$ and $\eta$, respectively. The energy spectrum
($-1/n^2$) and the degeneracy ($n^2$) of course do not depend on the
coordinate system. They are usually employed in the study of the
Stark effect as first shown by Epstein \cite{epstein}.

\medskip

\indent {\em Spheroidal}.

\noindent Spheroidal coordinates also can be treated by the
separation technique with the $z$ component of angular momentum
remaining as a constant of the motion. There are two types of
spheroidal coordinates.

The oblate spheroidal coordinates are given by:
\begin{eqnarray}\label{oblate}
x&=&r\cosh \xi \cos \eta \cos \phi ~,\nonumber \\
y&=&r\cosh \xi \cos \eta \sin \phi ~, \\
z&=&r\sinh \xi \sin \eta ~, \nonumber
\end{eqnarray}
where $\xi \in [0,\infty )$, $\eta \in [-\pi /2 , \pi /2]$, and
$\phi \in [0, 2\pi )$.

The prolate spheroidal coordinates are complementary to the oblate
ones in the variables $\xi$ and $\eta$:
\begin{eqnarray}\label{prolate}
x&=&r\sinh \xi \sin \eta \cos \phi ~,\nonumber \\
y&=&r\sinh \xi \sin \eta \sin \phi ~, \\
z&=&r\cosh \xi \cos \eta ~, \nonumber
\end{eqnarray}
where $\xi \in [0,\infty )$, $\eta \in [0 , \pi ]$, and $\phi \in
[0, 2\pi )$.

\medskip

\indent {\em Spheroconal}.

\noindent The spheroconal system is a quite unfamiliar coordinate
separation system of the H-atom Schr\"odinger equation. In this case
$\hat{L}^2$ is retained as a constant of the motion but
$\hat{L}_{z}^{2}$ is replaced by another separation operator
$\hat{B}$. The relationship with the Cartesian system is given
through \cite{cf}
\begin{eqnarray}\label{sconal1}
x&=& \frac{r}{bc}\,\theta \lambda \ , \nonumber \\
y&=& \frac{r}{bc}\left(-\frac{(b^2-\theta ^2)(b^2-\lambda
^2)}{1-\frac{b^2}{c^2}}\right)^{1/2} \ , \\
z&=& \frac{r}{bc}\left(-\frac{(c^2-\theta ^2)(c^2-\lambda
^2)}{1-\frac{b^2}{c^2}}\right)^{1/2}~. \nonumber
\end{eqnarray}
One of the separation constants is $l(l+1)$, the eigenvalue of the
separation operator $\hat{L}^2$. This fact is of considerable help
in dealing with the unfamiliar equations resulting from the
separation of the $\theta$ and $\lambda$ equations. The separation
operator $\hat{B}$ can be transformed into a Cartesian form,
\begin{equation}\label{23}
\hat{B}=b^2\hat{L}_y^2+ c^2\hat{L}_z^2~,
\end{equation}
where $\hat{L}_y$ and $\hat{L}_z$ are the usual Cartesian components
of the angular momentum operator. Thus, in the spheroconal system, a
linear combination of the squares of the $y$ and $z$ components of
the angular momentum effects the separation. The linear combination
coefficients are simply the squares of the limits of the spheroconal
coordinate ranges.\\

\medskip

In general, the stationary states in any of these coordinate systems
can be written as linear combination of degenerate eigenfunctions of
the other system and the ground state (the vacuum) should be the
same.

\bigskip

\section{\Large The 3D Parabolic Well: The Stationary States of the Isotropic Harmonic Oscillator}

We commented on the importance in physics of the HO at the beginning
of our analysis of the 1D quantum HO. If we will consider a 3D
analog, we would be led to study a Taylor expansion of the potential
in all three variables retaining the terms up to the second order,
which is a quadratic form of the most general form

\begin{equation}\label{3ho1}
V(x,y,z)=ax^2+by^2+cz^2+dxy+exz+fyz~.
\end{equation}

There are however many systems with spherical symmetry or for which
this symmetry is sufficiently exact. Then, the potential takes the
much simpler form
\begin{equation}\label{3ho2}
 V(x,y,z)=K(x^2+y^2+z^2)~.
\end{equation}

\noindent This is equivalent to assuming that the second unmixed
partial spatial derivatives of the potential have all the same
value, herein denoted by $K$. We can add up that this is a good
approximation whenever the values of the mixed second partial
derivatives are small in comparison to the unmixed ones. When these
conditions are satisfied and the potential is given by (\ref{3ho2}),
we say that the system is a 3D spherically symmetric HO.\\
The Hamiltonian in this case is of the form

\begin{equation}\label{3ho3}
\hat{H}=\frac{-\hbar^2}{2m}\bigtriangledown^2 +
\frac{m\omega^2}{2}r^2~,
\end{equation}
\noindent where the Laplace operator is given in spherical
coordinates and $r$ is the spherical radial coordinate.
Equivalently, the problem can be considered in Cartesian coordinates
but it is a trivial generalization
of the one-dimensional case.\\
Since the potential is time independent the energy is conserved. In
addition, because of the spherical symmetry the orbital momentum is
also conserved. Having two conserved quantities, we associate to
each of them a corresponding quantum number. As a matter of fact, as
we have seen in the case of the hydrogen atom, the spherical
symmetry leads to three quantum numbers, but the third one, the
magnetic number, is related to the `space quantization' and not to
the geometrical features of the motion.

Thus, the eigenfunctions depend effectively only on two quantum
numbers. The eigenvalue problem of interest is then

\begin{equation}\label{3ho4}
\hat{H}\Psi_{nl}=E_{nl}\Psi_{nl}~.
\end{equation}

The Laplace operator in spherical coordinates reads
\begin{equation}\label{3ho5}
\bigtriangledown^2 =\frac{\partial^2}{\partial
r^2}+\frac{2}{r}\frac{\partial}{\partial r}
-\frac{\hat{L}^2}{\hbar^2r^2} \ ,
\end{equation}
where the angular operator $\hat{L}^2$ is the usual spherical one
\begin{equation}\label{3ho6}
\hat{L}^2=-\hbar^2\left[ \frac{1}{\sin{\theta}}
\frac{\partial}{\partial\theta} \left(
\sin{\theta}\frac{\partial}{\partial\theta}\right)
+\frac{1}{\sin{\theta}^2}\frac{\partial^2}{\partial\varphi^2}\right]~.
\end{equation}

The eigenfunctions of $\hat{L}^2$ are the spherical harmonics, i.e.

\begin{equation}\label{3ho7}
\hat{L}^2Y_{lm_{l}}(\theta,\varphi)=-\hbar^2l(l+1)Y_{lm_{l}}(\theta,\varphi)
\ .
\end{equation}

In order to achieve the separation of the variables and functions,
the following substitution is proposed
\begin{equation}\label{3ho8}
\Psi_{nlm_{l}}(r, \theta,\varphi)=\frac{R_{nl}(r)}{r}
Y_{lm_{l}}(\theta,\varphi)~.
\end{equation}
Once this is plugged in the Schr\"odinger equation, the spatial and
the angular parts are separated from one another. The equation for
the spatial part has the form
\begin{equation}\label{3ho9}
 R_{nl}^{\prime\prime}+\left(\frac{2mE_{nl}}{\hbar^2}
-\frac{m^2\omega^2}{\hbar^2}r^2-\frac{l(l+1)}{r^2}\right)R_{nl}(r)=0~.
\end{equation}

Using the oscillator parameters $k_{nl}^{2}=\frac{2mE_{nl}}{\hbar
^2}$ and $\lambda=\frac{m\omega}{\hbar}$, the previous equation is
precisely of the one-dimensional quantum oscillator form but in the
radial variable and with an additional angular momentum barrier
term, 
\begin{equation}\label{3ho10}
R_{nl}^{\prime\prime}+\left(k_{nl}^{2}-\lambda^2r^2-\frac{l(l+1)}{r^2}\right)R_{nl}=0~.
\end{equation}

\noindent To solve this equation, we shall start with its asymptotic
analysis. Examining first the infinite limit $r\rightarrow\infty$,
we notice that the orbital momentum term is negligible and therefore
in this limit the asymptotic behavior is similar to that of the
one-dimensional oscillator, i.e., a Gaussian tail

\begin{equation}\label{3ho11}
R_{nl}(r)\sim\exp\left(-\frac{\lambda
r^2}{2}\right)\hspace{2cm}\mbox{for}
\hspace{.3cm}r\rightarrow\infty~.
\end{equation}

If now we consider the behavior near the origin, we can see that the
dominant term is that of the orbital momentum, i.e., the
differential equation (\ref{3ho10}) in this limit turns into

\begin{equation}\label{3ho12}
 R_{nl}^{\prime\prime}-\frac{l(l+1)}{r^2}R_{nl}=0~.
 \end{equation}

This is a differential equation of the Euler type
\[r^n y^{(n)}(r)+r^{n-1} y^{(n-1)}(r)+\cdots+r y^{\prime}(r)+y(r)=0\]
for the case $n=2$ with the first derivative missing. For such
equations the solutions are sought of the form $y=r^{\alpha}$ that
plugged in the equation lead to a simple polynomial equation in
$\alpha$, whose two independent solutions are $l+1$ and $-l$. Thus,
one gets

\begin{equation}\label{3ho13}
R_{nl}(r)\sim \hspace{.2cm}r^{l+1}\hspace{.2cm}\mbox{or}
\hspace{.4cm}r^{-l}\hspace{2cm}\mbox{for}\hspace{.4cm}r \rightarrow
0~.
\end{equation}

The previous arguments lead to proposing the substitution
\begin{equation}\label{3ho14}
R_{nl}(r)=r^{l+1}\exp{\left(-\frac{\lambda r^2}{2}\right)}u(r)~.
\end{equation}

The second possible substitution
\begin{equation}\label{3ho15}
R_{nl}(r)=r^{-l}\exp{\left(-\frac{\lambda r^2}{2}\right)}v(r)~,
\end{equation}
produces the same equation as (\ref{3ho14}).
Substituting (\ref{3ho14}) in (\ref{3ho12}), the following
differential equation for $u$ is obtained
\begin{equation}\label{3ho16}
u^{\prime\prime}+2\left(\frac{l+1}{r}-\lambda r\right)u^{\prime}
-[(2l+3)\lambda-k_{nl}^{2}]u=0~.
\end{equation}

By using now the change of variable $w=\lambda r^2$, one gets
\begin{equation}\label{3ho17}
wu^{\prime\prime}+\left(l+\frac{3}{2}-w\right)u^{\prime}-\left[
\frac{1}{2}\left(l+ \frac{3}{2}\right)-\frac{\kappa
_{nl}}{2}\right]u=0~,
\end{equation}
where $\kappa _{nl}=\frac
{k_{nl}^{2}}{2\lambda}=\frac{E_{nl}}{\hbar\omega}$ is the same
dimensionless energy parameter as in the one-dimensional case but
now with two subindices. We see that we found again a differential
equation of the confluent
hypergeometric type having the solution  

$$
u(r)=A\hspace{.2cm}_1F_1\left[\frac{1}{2}\left(l+\frac{3}{2}-\kappa
_{nl}\right),l+\frac{3}{2}; \lambda r^2\right]
$$
\begin{equation}\label{3ho18}
+B\hspace{.2cm}r^{-(2l+1)}
\hspace{.3cm}_1F_1\left[\frac{1}{2}\left(-l+\frac{1}{2}-\kappa
_{nl}\right),-l+\frac{1}{2}; \lambda r^2\right]~.
\end{equation}

The second particular solution cannot be normalized because it
diverges strongly for $r\rightarrow 0$. Thus one takes $B=0$ which
leads to

\begin{equation}\label{3ho19}
u_{B=0}(r)=A\hspace{.2cm}_1F_1\left[\frac{1}{2}\left(l+\frac{3}{2}-\kappa
_{nl}\right),l+\frac{3}{2}; \lambda r^2\right]~. \end{equation}

By using the same arguments on the asymptotic behavior as in the
one-dimensional HO case, that is, imposing a regular solution at
infinity, leads to the truncation of the confluent hypergeometric
series, which implies the quantization of the energy. The truncation
is explicitly
\begin{equation}\label{3ho20}
\frac{1}{2}\left(l+\frac{3}{2}-\kappa _{nl}\right)=-n~,
\end{equation}
\noindent and substituting $\kappa _{nl}$, we get the energy
spectrum
\begin{equation}\label{3ho21}
E_{nl}=\hbar\omega\left(2n+l+\frac{3}{2}\right)=\hbar\omega\left(N+\frac{3}{2}\right)
~.
\end{equation}
One can notice that for the three-dimensional spherically symmetric
HO there is a
zero point energy of $\frac{3}{2}\hbar\omega$, three times bigger than in the one-dimensional case.\\
The unnormalized eigenfunctions of the three-dimensional harmonic
oscillator are
\begin{equation}\label{3ho22}
\psi_{nlm_l}(r,\theta,\varphi) =r^{l}e^{\left(-\frac{\lambda
r^2}{2}\right)}\hspace{.2cm}_1F_1\left(-n,l +\frac{3}{2};\lambda
r^2\right)\hspace{.1cm}Y_{lm_l}(\theta,\varphi)~.
\end{equation}
Since we are in the case of a radial problem with centrifugal
barrier we know that we have to get as solutions the associated
Laguerre polynomials. This can be seen by using the condition
(\ref{3ho20}) in (\ref{3ho17}) which becomes
\begin{equation}\label{3ho23}
wu^{\prime\prime}+\left(l+\frac{3}{2}-w\right)u^{\prime}+nu=0~.
\end{equation}
The latter equation has the form of the associated Laguerre equation
$wu^{\prime\prime}+\left(p+1-w\right)u^{\prime}+nu=0$ for $p=l+1/2$
with the polynomial solutions $L_{n}^{l+1/2}$. Thus, the normalized
solutions can be written
\begin{equation}\label{3ho24}
\psi_{nlm_l}(r,\theta,\varphi) ={\cal
N}_{nl}r^{l}e^{\left(-\frac{\lambda
r^2}{2}\right)}\hspace{.2cm}L_{n}^{l+1/2}(\lambda
r^2)\hspace{.1cm}Y_{lm_l}(\theta,\varphi)~,
\end{equation}
where the normalization integral can be calculated as in the 1D
oscillator case using the Laguerre generating function
\begin{equation}\label{Laggen}
\frac{e^{-zt/(1-t)}}{(1-t)^{k+1}}=\sum
_{p=0}^{\infty}\frac{t^p}{p+k)!}L_p^k(z)~.
\end{equation}
The final result is
\begin{equation}\label{normlagosc}
{\cal N}_{nl}=
\bigg[\sqrt{\frac{\lambda}{\pi}}\frac{n!(n+l)!}{2^{-(2n+2l+2)}(2n+2l+1)!}\bigg]^{\frac{1}{2}}~.
\end{equation}
If the dynamical phase factor ${\cal F}$ is included, the stationary
wavefunctions of the 3D spherically symmetric oscillator take the
following final form
\begin{equation} \label{osc3wfin}
\psi_{n,l,m_l} (r,\theta, \phi, t)={\cal N}_{nl} \exp
\left(-i\left(2n+l+\frac{3}{2}\right)\omega t-\frac{\lambda
r^2}{2}\right)  r^{l}\,L_{n}^{l+1/2}(\lambda
r^2)\hspace{.1cm}Y_{lm_l}(\theta,\varphi)~.
\end{equation}
For the algebraic (factorization) method applied to the radial
oscillator we refer the reader to detailed studies \cite{mota03}.
The degeneracy of the radial oscillator is easier to calculate by
counting the Cartesian eigenstates at a given energy, which gives
$(N+2)!/N!2!$ \cite{sa96}.

\bigskip

\section{Stationary Bound States in the Continuum}

After all these examples, it seems that potential wells are
necessary for the existence of stationary states in wave mechanics.
However, this is not so! All of the quantum bound states considered
so far have the property that the total energy of the state is less
than the value of the potential energy at infinity, which is similar
to the bound states in classical mechanics. The boundness of the
quantum system is due to the lack of sufficient energy to
dissociate. However, in wave mechanics it is possible to have bound
states that do not possess this property, and which therefore have
no classical analog.

\medskip

Let us choose the zero of energy so that the potential energy
function vanishes at infinity. The usual energy spectrum for such a
potential would be a positive energy continuum of unbound states,
with the bound states, if any, occurring at discrete negative
energies. However, Stillinger and Herrick (1975) \cite{ball1},
following an earlier suggestion by Von Neumann and Wigner
\cite{ball2}, have constructed potentials that have discrete bound
states embedded in the positive energy continuum. Bound states are
represented by those solutions of the equation $(-\frac{1}{2}\nabla
^2+V)\Psi =E\Psi$ for which the normalization integral $\int |\Psi
|^2d^3x$ is finite. (We adopt units such that $\hbar =1$ and $m$
=1.)

\medskip

We can formally solve for the potential,
\begin{equation}\label{ball-1}
V(r;E)=E+\frac{1}{2}\left(\frac{\nabla ^2 \Psi}{\Psi}\right)~.
\end{equation}
For the potential to be nonsingular, the nodes of $\Psi$ must be
matched by zeros of $\nabla ^2 \Psi$. The free particle
zero-angular-momentum function $\Psi _0({\bf x})=\sin(kr)/kr$
satisfies (\ref{ball-1}) with energy eigenvalue $E_0=\frac{1}{2}k^2$
and with $V$ identically equal to zero, but it is unacceptable
because the integral of $|\Psi _0|^2$ is not convergent. However, by
taking
\begin{equation}\label{ball-2}
\Psi({\bf x})=\Psi _0({\bf x}) f(r) \ ,
\end{equation}
and requiring that $f(r)$ go to zero more rapidly than $r^{-1/2}$ as
$r\rightarrow \infty$ one can get a convergent integral for
$|\Psi(({\bf x})|^2$. Substituting (\ref{ball-2}) into
(\ref{ball-1}), we obtain
\begin{equation}\label{ball-3}
V(r;E)=E-\frac{1}{2}k^2+k\cot(kr)\frac{f'(r)}{f(r)}+\frac{1}{2}\frac{f''(r)}{f(r)}~.
\end{equation}
For $V$ to remain bounded, $f'(r)/f(r)$ must vanish at the poles of
$\cot (kr)$; that is, at the zeros of $\sin (kr)$. This can be
achieved in different ways but in all known procedures the
modulation factor $f(r)$ has the form
\begin{equation}\label{ball-5}
f(r)=[\lambda +s(r)]^{-1}~,
\end{equation}
where $\lambda$ is a positive constant, although Stillinger and
Herrick mentioned a wider class of possible $f(r)$. They chose the
modulation variable
\begin{equation}\label{ball-4}
s_{sh}(r)=8k^2\int _0^r
r'[\sin(kr')]^2dr'=\frac{1}{2}(2kr)^2-2kr\sin(2kr)-\cos(2kr) +1~.
\end{equation}
The principles guiding the choice of $s(r)$ are: that the integrand
must be nonnegative, so that $s(r)$ will be a monotonic function of
$r$; and that the integrand must be proportional to $\sin (kr)$, so
that $ds(r)/dr$ will vanish at the zeros of $\sin(kr)$.
For $s_{sh}$, $\Psi$ decreases like $r^{-3}$ as $r\rightarrow
\infty$, which ensures its square integrability. The potential
(\ref{ball-3}) then becomes (for $E_0$)
\begin{equation}\label{ball-6}
V_{sh}(r;E_0)=\frac{64k^4r^2[\sin(kr)]^4}{[\lambda
+s_{sh}(r)]^2}-\frac{4k^2\{\,[\sin (kr)]^2+2kr\sin(2kr)\}}{\lambda
+s_{sh}(r)}~.
\end{equation}

\medskip

The energy of the bound state produced by the potential $V(r;E_0)$
is $E_0=\frac{1}{2}k^2$ as for the free particle, i.e., it is
independent of $\lambda$ and the modulation factor $f(r)$.
Therefore, the main idea for getting bound states in the continuum
is to build isospectral potentials of the free particle and more
generally for any type of scattering state. A more consistent
procedure to obtain isospectral potentials is given by the formalism
of supersymmetric quantum mechanics \cite{psp93} which is based on
the Darboux transformations \cite{d1882}. For the supersymmetric
case the modulation variable is
\begin{equation}\label{ball-8}
s_{d}(r)=\int _0^r u_0^2 (r')dr'=\int _0^r
[\sin(kr')]^2dr'=\frac{1}{2}r-\frac{1}{4k}\sin(2kr)
\end{equation}
and the isospectral potential has the form
\begin{equation}\label{ball-9}
V_d(r;E_0)=\frac{2[\sin(kr)]^4}{[\lambda
+s_{d}(r)]^2}-\frac{2k\,\sin(2kr)}{\lambda +s_{d}(r)}~.
\end{equation}
Finally, in the amplitude modulation method of von Neumann and
Wigner the modulation variable is given by
\begin{equation}\label{ball-10}
s_{vnw}(r)=(4ks_d)^2=[2kr-\sin(2kr)]^2
\end{equation}
and the isospectral free particle potential is
\begin{equation}\label{ball-11}
V_{vnw}(r;E_0)=-\frac{64k^2\lambda [\sin(kr)]^4}{[\lambda
+s_{vnw}(r)]^2}+\frac{48k^2[\sin
(kr)]^4-8k^2s_{vnw}^{1/2}\sin(2kr)}{\lambda +s_{vnw}(r)}~.
\end{equation}
Interestingly, all these potentials have the same behavior at large
$r$
\begin{equation}\label{ball-7}
V_{sh}(r;E_0)\sim V_{d}(r;E_0)\sim V_{vnw}(r;E_0)\approx -\frac{4k
\sin(2kr)}{r}=-8k^2{\rm sinc}(2kr)=-8k^2j_0(2kr)~,
\end{equation}
where the sinc is the cardinal sine function which is identical to
the spherical Bessel function of the first kind $j_0$. On the other
hand, these potentials display different power-law behavior near the
origin.

Moreover, the asymptotic sinc form which is typical in diffraction
suggests that the existence of bound states in the continuum can be
understood by using the analogy of wave propagation to describe the
dynamics of quantum states. It seems that the mechanism which
prevents the bound state from dispersing like ordinary positive
energy states is the destructive interference of the waves reflected
from the oscillations of $V(r;E)$. According to Stillinger and
Herrick no other $f(r)$ that produces a single particle bound state
in the continuum will lead to a potential that decays more rapidly
than (\ref{ball-7}). However, they present further details in their
paper which suggest that nonseparable multiparticle systems, such as
two-electron atoms, may possess bounded states in the continuum
without such a contrived form of potential as (\ref{ball-6}).

The bound (or localized) states in the continuum are not only of
academic interest. Such a type of electronic stationary quantum
state has been put into evidence in 1992 by Capasso and
collaborators \cite{cap92} by infrared absorption measurements on a
semiconductor superlattice grown by molecular beam epitaxy in such a
way that one thick quantum well is surrounded on both sides by
several GaInAs-AlInAs well/barrier layers constructed to act as
$\lambda/4$ Bragg reflectors. There is currently much interest in
such states in solid state physics \cite{ssp}.

\section{Conclusion}

We have reviewed the concepts that gave rise to quantum mechanics.
The stationary states were introduced first by Bohr to explain the
stability of atoms and the experimental findings on the variation of
the electronic current when electrons collided with mercury atoms.
The generalization of those ideas were discussed and the
Schr\"odinger equation was introduced. The stationary localized
solutions of that equation for various potentials with closed and
open boundary conditions were worked out in detail and the physical
meaning was stressed.

There are other cases of interest, like the solution of the
Schr\"oedinger equation of electrons moving in a solid, that are
treated in detail in other chapters. Our interest here is to stress
the historic development of quantum mechanics and to show the
importance of the stationary state concept.

\end{document}